\documentclass[
reprint,
nofootinbib,
longbibliography,
amsmath,amssymb,amsfonts
 aps,
onecolumn,
prd
]{revtex4-1}

\usepackage{natbib}

\usepackage{xcolor}

\usepackage{bbold}
\usepackage{dsfont}
\usepackage{nccmath}

\usepackage{setspace}
\usepackage[fontsize=10]{scrextend}

\DeclareFontFamily{OT1}{pzc}{}
\DeclareFontShape{OT1}{pzc}{m}{it}{<-> s * [1.10] pzcmi7t}{}
\DeclareMathAlphabet{\mathpzc}{OT1}{pzc}{m}{it}

\usepackage[titletoc,toc,title]{appendix}
\usepackage{graphicx}% Include figure files
\usepackage{dcolumn}% Align table columns on decimal point
\usepackage{bm}% bold math
\usepackage[utf8]{inputenc}
\usepackage{amsmath}
\usepackage{amsfonts}
\usepackage{mathtools}
\usepackage{dsfont}
\usepackage{mathrsfs}
\usepackage{float} % Allows putting an [H] in \begin{figure} to specify the exact location of the figure
\usepackage{wrapfig} % Allows in-line images such as the example fish picture
\usepackage[caption=false]{subfig}
\usepackage{geometry}
\usepackage[colorlinks,allcolors=blue,bookmarks=false,hypertexnames=true]{hyperref} 

\usepackage{enumitem}

\usepackage{titlesec}
\titleformat{\section}{\centering\bfseries\fontsize{10pt}{10}\selectfont}{\Roman{section}.}{1em}{}
\titleformat{\subsection}{\centering\bfseries\fontsize{10pt}{10}\selectfont}{\Alph{subsection}.}{1em}{}

\usepackage{etoolbox}

\begin{document}

%\preprint{APS/123-QED}

\title{Dark D-Brane Cosmology: from background evolution to cosmological perturbations}% Force line breaks with \\

\author{Carsten van de Bruck and Elsa M. Teixeira}
\affiliation{%
School of Mathematics and Statistics, University of Sheffield, Hounsfield Road, Sheffield S3 7RH, United Kingdom
}%

\begin{abstract}
We study the cosmological predictions of the dark D--brane model, in which dark matter resides on a D--brane moving in a higher-dimensional space. By construction, dark matter interacts only gravitationally with the standard model sector in this framework. The dark energy scalar field is associated with the position of the D--brane, and its dynamics is encoded in a Dirac--Born--Infeld action. On the other hand, dark matter is identified with matter on the D-brane, that naturally couples to dark energy \textit{via} a disformal coupling. We analyse the numerical evolution of the cosmological background, highlighting the fact that there are two regimes of interest: one, in which the coupling is positive throughout and another, in which the coupling is negative at the present. In the latter, there is the enticing possibility of having scenarios in which the coupling is positive for a significant part of the evolution, before decreasing towards negative values. In both cases, the coupling is very small at early times, and only starts to grow during the late matter dominated era. We also derive the equations for the linear cosmological perturbations, an expression for the effective time-dependent gravitational coupling between dark matter particles and present the numerical results for the cosmic microwave background anisotropy and matter power spectra. This allows for a direct comparison of the predictions for the growth of large scale structure with other disformal quintessence models.

%\begin{description}
%\end{description}
\end{abstract}

\pacs{Valid PACS appear here}% PACS, the Physics and Astronomy
\maketitle
\singlespacing

%\tableofcontents

\section{\label{sec:int}Introduction}
In the past few decades, cosmologists have developed a framework which explains many properties of the observable Universe \cite{1804633}. Its successes rely on the existence of a dark sector, whose origin most likely comes from new physics beyond the standard model of particle physics. In the simplest version, the dark sector consists of two parts \cite{Maddox:1990hb,Efstathiou:1990xe,Ostriker:1995su}. The first is some form of nonrelativistic particle, cold dark matter (CDM), which interacts only very weakly, if at all, with the standard model fields. Dark matter is needed to explain a plethora of observations, such as the rotation curves of galaxies and the motion of galaxies in galaxy clusters \cite{Bertone:2004pz,Clowe:2006eq}. Moreover, it also provides an explanation of how structures form and makes predictions about the statistical properties of anisotropies in the cosmic microwave background (CMB), which have been confirmed by experiments. The second part is motivated by the observed accelerated expansion of the Universe \cite{300499,9812133}, a phenomenon generally attributed to a new energy form dubbed dark energy (DE). In the simplest scenario, dark energy is portrayed by the cosmological constant \cite{Carroll:2000fy,Peebles:2002gy} and the dark sector remains uncoupled.

While this $\Lambda$CDM model remains the most economic scenario for explaining a variety of cosmological observations, the physical origin of the dark sector remains unclear. It is generally believed that CDM is a particle, predicted by most extensions of the standard model. On the other hand, the magnitude of the cosmological constant is difficult to explain within conventional quantum field theories \cite{Rugh:2000ji,Weinberg:2000yb,Martin:2012bt}. This motivated the formulation of alternative frameworks, primarily with the introduction of the quintessence canonical scalar field, whose dynamical evolution can resemble a cosmological-constant-like behaviour \cite{Wetterich:1987fm,PhysRevD.37.3406,AstrophysJ.325.L17,Caldwell:1997ii,Tsujikawa:2013fta}.  Not many works address the possibility of a joint origin for dark matter and dark energy (see \textit{e.g.} \cite{Gao:2009me,Ansoldi:2012pi,Arbey:2020ldf,Brandenberger:2018xnf}), but there are a plethora of models proposed in the literature in the dark sector is coupled, seminally proposed and studied in Refs. \cite{ELLIS1989264,Wetterich:1994bg,Amendola:1999er,Holden:1999hm} (for a general overview see \cite{Copeland:2006wr,Amendola:2015ksp}). In this work we will study a model in which dark matter and dark energy, while still being two distinctive components, have a joint {\it higher-dimensional origin} \cite{Koivisto:2013fta}. Specifically, the dark matter sector originates from matter on a D--brane, moving in a higher-dimensional spacetime. The role of the dark energy field is played by the position of the brane, whose motion in the extra dimensions is encoded in kinetic and potential energy terms in the low-energy effective action. This D--brane scenario arises from hidden sector branes in String Theory, which have no intersection with D--branes responsible for the visible (standard model) sector, and, therefore, dark matter interacts only \textit{via} gravity with the standard model fields (for a review on cosmological applications of String Theory see, \textit{e.g.}, \cite{McAllister:2007bg}). Although dark energy and dark matter are still two separate components, in this scenario they both stem from properties of the D--brane, hence being inevitably coupled. Furthermore, because the object from which dark energy originates is a D--brane, the kinetic term is noncanonical and takes the form of a Dirac--Born--Infeld (DBI) kinetic term, widely studied in cosmology, namely, in the context of inflation \cite{
%Garriga:1999vw,
Silverstein:2003hf,Alishahiha:2004eh,
%Panda:2005sg,
Chimento:2007es} and dark energy models \cite{Abramo:2004ji,Martin:2008xw,Guo:2008sz,Gumjudpai:2009uy,Ahn:2009hu,Ahn:2009xd,Chimento_2009,Copeland:2010jt,Brax:2012jr,Kaeonikhom:2012xr,Burrage:2014uwa,Mahata:2015lja,Panpanich:2017nft}.

Dark matter and dark energy are naturally coupled in this scenario, because the metrics which define geodesics for standard model particles and dark matter are not the same. Therefore, the coupling of the scalar field to matter is nonuniversal by construction, avoiding conflicts with constraints from Solar System tests \cite{Will:2014kxa,Sakstein:2014isa,Ip:2015qsa,Wang:2016lxa} and with the strict bounds on the speed of gravitational waves \cite{Monitor:2017mdv}.
 The relation between the two metrics is given by a disformal transformation \cite{Bekenstein:1992pj}, as will be explained in Sec. \ref{sec:model} in more detail, that takes the form

\begin{equation}
\bar{g}_{\mu \nu} = C \left( \phi\right) g_{\mu \nu} + D \left( \phi \right) \partial_{\mu} \phi \partial_{\nu} \phi.
\label{trans}
\end{equation}

\noindent The conformal factor $C(\phi)$ and the disformal factor $D(\phi)$ are related to each other and carry information about the curvature of the extradimensional space. 
%The disformal relation gained special attention in the cosmology community when it was shown that the Horndeski Lagrangian \cite{Horndeski:1974wa} is invariant under disformal transformations of the form of Eq. \eqref{trans}, \cite{Zumalacarregui:2013pma,Bettoni:2013diz}. 
The disformal relation has been previously proposed in brane world cosmological models \cite{KOIVISTO:2013jwa,Koivisto:2014gia,Koivisto:2015vda,Cembranos:2016jun} and has been applied to several different areas of study in cosmology \cite{deRham:2010ik,Clayton:1998hv,Bettoni:2011fs,Deruelle:2014zza,Brax:2012hm,Olmo:2009xy,Sakstein:2014aca,Ezquiaga:2017ner}. It has also been used to study inflationary settings \cite{Kaloper:2003yf,vandeBruck:2015tna}, disformal quintessence fields \cite{Koivisto:2008ak,Zumalacarregui:2010wj}, and models of disformally coupled dark energy \cite{Noller:2012sv,Bettoni:2012xv,Koivisto:2012za,Zumalacarregui:2012us,Zumalacarregui:2013pma,Bettoni:2013diz,Sakstein:2014aca,Sakstein:2015jca,vandeBruck:2015ida,Bettoni:2015wla,vandeBruck:2016jgg,Teixeira:2019hil}.

In this paper, we further investigate the model proposed in Ref. \cite{Koivisto:2013fta}, where the cosmological background evolution was discussed by means of a dynamical systems analysis. In addition to a numerical study of the background evolution, taking into account radiation and baryonic matter, we also discuss for the first time aspects of cosmological perturbations in this theory. We present the perturbation equations in both the Newtonian and synchronous gauges. We solve the full set of cosmological equations numerically and present predictions such as the CMB temperature anisotropies and matter power spectra. We will furthermore compare features of the dark D--brane cosmological scenario to other dark energy theories with disformal couplings, such as the ones proposed in Refs. \cite{Zumalacarregui:2012us,vandeBruck:2015ida,vandeBruck:2017idm,Mifsud:2017fsy,Teixeira:2019hil}. To search for new signatures predicted in coupled dark energy models, an in-depth study of the growth of perturbations is of paramount importance\footnote{See \textit{e.g.} \cite{Bettoni:2012xv,Pace:2013pea,Minamitsuji:2014waa,Tsujikawa:2014uza,vandeBruck:2015ida,Barros:2018efl,Frusciante:2020zfs} for previous work on diverse models with a dark energy-dark matter coupling.}.

The paper is organised as follows. In Sec. \ref{sec:model} we present the details of the theory and write down the general equations of motion. In Sec. \ref{sec:back} we discuss the cosmological evolution, focusing on the case of a flat Friedmann--Lema\^{i}tre--Robertson--Walker (FLRW) spacetime. We turn our attention to cosmological perturbations in Sec. \ref{sec:pert}, where we present the general equations in the Newtonian gauge (the equations in the synchronous gauge are given in Appendix \ref{app:synch}). We derive an expression for the effective gravitational constant between dark matter particles and show that the gravitational constant grows at late times in the dark D--brane cosmological model. We also present results for the CMB anisotropy and matter power spectra and discuss the evolution of the density contrast of dark matter. A summary of our findings and conclusions can be found in Sec. \ref{sec:conc}.

\section{The Model} \label{sec:model}

In the model we consider in this work, the dark sector originates from a hidden D3-brane, moving in a higher-dimensional spacetime, comprised of two degrees of freedom: some kind of matter fields confined to the brane and the brane's radial position\footnote{In this work we consider the case of only one species living on the hidden brane, which we assume is pressureless so that it acts as cold dark matter.}. Cold dark matter is identified with particles living on the D3-brane, which has no intersection with the D--branes from which the standard model fields originate (hence, the D--brane is hidden). For this reason, dark matter and the standard model fields interact only gravitationally in the low-energy field theory. The role of dark energy is played by the scalar representing the position of the brane in the extra dimensions, leading to an interaction in the dark sector. The geometry of the higher-dimensional space is encoded in the warp factor, which we assume depends on the radial coordinate only, and, further on, we will focus on AdS$_5\times$S$^5$ warped regions. 

The resulting theory we consider was constructed in Ref. \cite{Koivisto:2013fta} by considering a warped flux compactification of type IIB String Theory. The low-energy 4D effective action is of the form (we adopt conventions of $c= \hbar = 1$ and metric signature $-+++$)

\begin{eqnarray}
\label{action}
S & = &\frac{1}{2 \kappa^2} \int d^4 x \sqrt{-g} R\, + \int d^4 x \sqrt{-g} \left[h^{-1}(\phi) \left( 1 - \sqrt{1 + h(\phi) \partial^{\mu} \phi \partial_{\mu} \phi} \right) - V(\phi) \right] \\
& + & \sum_i \int d^4 x \sqrt{-g} \mathcal{L}_{S} \left( g_{\mu \nu}, \psi_i,  \partial_{\mu} \psi_i\right) + \sum_j \int d^4 x \sqrt{-\bar{g}} \mathcal{L}_{DDM} \left( \bar{g}_{\mu \nu}, \chi_j,  \partial_{\mu} \chi_j \right). \nonumber
\end{eqnarray} 

\noindent The first term is the standard Einstein-Hilbert action, where $\kappa^2 = M_{Pl}^{-2} = 8 \pi G_N$ is the reduced Planck mass, $G_N$ is Newton’s gravitational constant, $g$ is the determinant of the metric tensor $g_{\mu \nu}$, and $R$ is the Ricci scalar. The metric $g_{\mu\nu}$ is the metric which defines geodesics for the standard model fields. The second term corresponds to the scalar Dirac-Born-Infeld action \cite{Silverstein:2003hf,Alishahiha:2004eh} for a D3-brane, with the scalar field $\phi$ representing a canonical normalisation of the radial position $r$ of the D3--brane: $\phi \equiv \sqrt{T_3} r$, where $T_3$ is the tension of the brane. The warp factor also becomes field dependent, $h(\phi) \equiv T_3^{-1} h(r)$, carrying the geometrical information about the warped throat region in the compactified space. The term $V(\phi)$ is a potential function. The third and fourth terms are the actions for standard model particles, represented by the matter fields $\psi_i$, and for matter fields living on the D3--brane, $\chi_j$, respectively. In the latter, the fields propagate on geodesics specified by the induced metric on the brane $\bar{g}_{\mu \nu}$, shown to be related to the metric $g_{\mu\nu}$ by a disformal transformation of the form in Eq. \eqref{trans}. In this paper, we will assume that the matter on the D--brane is disformal cold dark matter (DDM). Because dark matter propagates on the D--brane, it is naturally coupled to the scalar field $\phi$.

Before we discuss cosmological applications of the action above, we remind the reader that some assumptions were made when deriving the low-energy effective action \cite{Koivisto:2013fta}. The most important one is to assume that the dark D-brane can be treated as a {\it probe brane}, i.e., implying that its presence does not backreact onto the background geometry. This means that the extra degrees of freedom that could potentially emerge due to the presence of the brane can, therefore, be ignored in this context. We will further on discuss cosmological perturbations in the model, which are small and treated at a linear level in perturbation theory only. Furthermore, the brane will itself never be in the highly relativistic regime, so that any backreaction on the bulk geometry due to this effect will also always remain negligible. Thus, for the purpose of this study, the action in Eq. \eqref{action} consists of a satisfactory framework to describe disformally coupled dark matter in this model. However, one should keep in mind that, when studying cosmological perturbations at the non linear level, further corrections to the action in Eq. \eqref{action} may have to be considered, emerging from additional degrees of freedom in the theory. 

As we will discuss below, the conformal and disformal functions are not independent but related to the warp factor $h(\phi)$. However, for completeness, we will write down the equations for general $C(\phi)$ and $D(\phi)$. The relation between the metrics encodes the phenomenological coupling between the DBI scalar field and CDM. As was previously mentioned, the standard model particles do not couple to the dark sector in this framework.

From the action \eqref{action}, we can derive the equations of motion. Einstein's equations read

\begin{equation}
G_{\mu \nu} \equiv R_{\mu \nu} - \frac{1}{2} g_{\mu \nu} R = \kappa^2 \left( T_{\mu \nu}^{\phi} + T_{\mu \nu}^{c} + T_{\mu\nu}^S \right),
\label{eins}
\end{equation}

\noindent where we have defined each energy-momentum tensor as 

\begin{equation}
T_{\mu \nu}^{\phi} = - \frac{2}{\sqrt{-g}} \frac{\delta \left( \sqrt{- g} \mathcal{L}_{\phi} \right)}{\delta g^{\mu \nu}}, \ \ 
T_{\mu \nu}^{c} = - \frac{2}{\sqrt{-g}} \frac{\delta \left( \sqrt{- \bar{g}} \mathcal{L}_{DDM} \right)}{\delta g^{\mu \nu}}, \ \ T_{\mu \nu}^{S} = - \frac{2}{\sqrt{-g}} \frac{\delta \left( \sqrt{- g} \mathcal{L}_{S} \right)}{\delta g^{\mu \nu}}.
\label{tmunu}
\end{equation}

\noindent Because of the presence of the coupling in the dark sector, the energy-momentum tensors of the scalar field and dark matter are not individually divergenceless. Still, in order to preserve general covariance, the total energy-momentum tensor must still be conserved through the Bianchi identities, given by 

\begin{equation}
\nabla_{\mu} \left(  T^{\mu \nu}_{\phi} + T^{\mu \nu}_{c} \right) = 0,\ \ \nabla_{\mu} T^{\mu \nu}_{S} =0.
\label{emtcons}
\end{equation}

From the action in Eq. \eqref{action}, we find the equation of motion for the scalar field, which reads 

\begin{equation}
\nabla_{\mu} \left( \gamma \partial^{\mu} \phi \right) - V_{, \phi} + \frac{h_{, \phi}}{h^2} \frac{\gamma}{2} \left( \gamma^{-1} -1 \right)^2 = \nabla_{\mu} \left[ \frac{D}{C} T_{c}^{\mu \alpha} \partial_{\alpha} \phi \right] - \frac{1}{2} \left[ \frac{C_{, \phi}}{C} T_{\rm c} + \frac{D_{, \phi}}{C} T_{\rm c}^{\mu \nu} \partial_{\mu} \phi \partial_{\nu} \phi \right],
\label{eqmogen}
\end{equation}

\noindent where the subscript ${ \phi}$ stands for derivatives with respect to the scalar field, $T_{\rm c} \equiv g_{\mu \nu} T_{\rm c}^{\mu \nu}$ is the trace of the dark matter energy-momentum tensor, and 

\begin{equation}
\gamma = \frac{1}{\sqrt{1+h(\phi) \partial^{\mu} \phi \partial_{\mu} \phi}},
\end{equation}

\noindent is called the Lorentz factor for the brane's motion, which must always be real. This term is a measure of the relativistic motion of the D3--brane. In the limit where $\gamma \rightarrow 1$ we recover the canonical kinetic term and, in this regime, $\phi$ behaves like a standard quintessence field. For $\gamma \rightarrow \infty$ we approach the purely relativistic limit.

So far all the equations were derived considering general conformal and disformal functions. We now focus on the case where the disformal metric in Eq. \eqref{trans} corresponds to the induced metric on a probe D3--brane moving in a warped higher dimensional spacetime. In this framework, the terms $C$ and $D$ in the transformation become functions of the warp factor of the brane, $h(\phi)$ \cite{Koivisto:2013fta}: 

\begin{equation}
C(\phi) = \left[ T_3 h(\phi) \right]^{-1/2}\ \ \text{and}\ \ D(\phi) = \left[ h(\phi)/T_3 \right]^{1/2}.
\label{transh}
\end{equation}

\noindent In this case, Eq. \eqref{eqmogen} reads 

\begin{equation}
\nabla_{\mu} \left( \gamma \partial^{\mu} \phi \right) - V_{, \phi} + \frac{h_{, \phi}}{2h^2} \gamma \left( \gamma^{-1} -1 \right)^2 = \nabla_{\mu} \left[ h(\phi) T_{c}^{\mu \nu} \partial_{\nu} \phi \right] - \frac{T_{c}^{\mu \nu}}{4} \left[ -\frac{h_{, \phi}}{h} g_{\mu \nu} + h_{, \phi} \partial_{\mu} \phi \partial_{\nu} \phi \right].
\end{equation}

Under the assumption that the Universe is homogeneous and isotropic, all the matter species in the theory can be modelled as perfect fluids.
For cold dark matter, this means that its corresponding energy-momentum tensor can be written as 

\begin{equation}
T^{c}_{\mu \nu} = \rho_c u^c_{\mu} u^c_{\nu},
\label{tmunuddm}
\end{equation}

\noindent where $u^c_{\mu}$ is the fluid's four velocity for a comoving observer and $\rho_c$ its energy density.

From the definition of the energy-momentum tensor for the scalar field, Eq. \eqref{tmunu}, and the DBI action, Eq. \eqref{action}, we compute

\begin{equation}
T_{\mu \nu}^{\phi} = \left( \frac{1-\gamma^{-1}}{h} - V \right) g_{\mu \nu} + \gamma \partial_{\mu} \phi \partial_{\nu} \phi.
\label{tmunudef}
\end{equation}

 \noindent Assuming a perfect fluid form for the dark energy fluid as well, we have

\begin{equation}
T_{\mu \nu}^{\phi} = p_{\phi} g_{\mu \nu} + \left( \rho_{\phi} + p_{\phi} \right) u_{\mu}^{\phi} u_{\nu}^{\phi}.
\label{tmunuperf}
\end{equation}

\noindent Comparing Eqs. \eqref{tmunudef} and \eqref{tmunuperf} yields

\begin{equation}
u_{\mu}^{\phi} = \frac{\partial_{\mu} \phi}{\sqrt{-\partial_{\nu} \phi \partial^{\nu} \phi}}
\end{equation}

\noindent for the scalar field's four velocity and

\begin{equation}
\rho_{\phi} = \frac{\gamma -1}{h} + V\ \ \text{and}\ \  p_{\phi} = \frac{1 - \gamma^{-1}}{h} - V
\label{rhopphi}
\end{equation}

\noindent for the energy density and pressure, respectively. We also define the equation of state (EoS) parameter of the scalar field as 

\begin{equation}
w_{\phi} = \frac{p_{\phi}}{\rho_{\phi}} = \frac{\left(\gamma-1 \right)/h\gamma - V }{\left(\gamma-1 \right)/h+ V}.
\label{wphi}
\end{equation}
The conservation relations in Eq. \eqref{emtcons} can be manipulated to give an analytic expression for the coupling between the dark fluids, which we denote by $Q$:

\begin{equation}
\nabla_{\mu} T^{\mu \nu}_{\phi} =  \left[ \nabla_{\mu} \left( \gamma \partial^{\mu} \phi \right) - V_{, \phi} + \frac{h_{, \phi}}{2h^2} \gamma \left( \gamma^{-1} -1 \right)^2 \right] \partial^{\nu} \phi = Q \partial^{\nu} \phi,
\label{cons}
\end{equation}
with 
\begin{equation}
Q= \nabla_{\alpha} \left[ h(\phi) T_{c}^{\alpha \beta} \partial_{\beta} \phi \right] - \frac{h_{, \phi}}{4 h} T_{c}^{\alpha \beta} \left[ - g_{\alpha \beta} + h \partial_{\alpha} \phi \partial_{\beta} \phi \right].
\label{Qcov}
\end{equation}

\noindent The interaction term $Q$ encodes the energy flows between the dark energy component and the dark matter sector and will play a fundamental role in this work.

\section{Background FLRW Cosmology} \label{sec:back}

We now turn our attention to the cosmological implications of the dark D--brane model. We specify the background spacetime to be a spatially flat FLRW Universe with line element 

\begin{equation}
ds^2 = a^2(\tau) \left( -d \tau^2 + dx^2 + dy^2 + dz^2 \right),
\end{equation}

\noindent where $\tau$ is the conformal time and $a(\tau)$ is the scale factor of the Universe. The scalar field is assumed to be homogeneous, that is, $\phi = \phi (\tau)$ is a function of time only and hereafter we will use primes and upper dots to denote derivatives with respect to conformal time $\tau$ and cosmic time $t$, respectively, related by $dt=a d\tau$. In the disformal frame, in which the dark matter geodesics are defined, the line element becomes

\begin{equation}
d\bar{s}^2 = C a^2(\tau) \left( - Z^2 d \tau^2 +  dx^2 + dy^2 + dz^2  \right),
\label{dst}
\end{equation}

\noindent where $Z$ stands for the disformal scalar, related to the Jacobian of the metric transformation. For the scenario considered in this work, with $C$ and $D$ as defined in Eq. \eqref{transh}, the disformal scalar be identified with the inverse of the Lorentz factor, \textit{i.e.} 

\begin{equation}
    Z \equiv \sqrt{1 - 2 X \frac{D}{C}} = \sqrt{1-h(\phi) \frac{\phi'^2}{a^2}} = \frac{1}{\gamma},
    \label{zdef}
\end{equation}

\noindent where $X$ stands for the standard kinetic term of the scalar field, $X=-\frac{1}{2}g^{\mu\nu} \partial_\mu\phi \partial_\nu \phi = \frac{1}{2a^2} \phi'^2$, and from where it is clear that $Z \geq 0$, with equality holding in the limit $\gamma \rightarrow \infty$ only. From Eq. \eqref{dst}, we can distinguish clearly the effect of the conformal and the disformal terms independently: While the conformal factor acts on the whole line element, modifying the expansion and thereby diluting dark matter over space and time, the disformal factor acts only on the time component, changing dark matter particles’ light cones.

In this context, with $T^{\mu \nu}_{c}$ as defined in Eq. \eqref{tmunuddm}, we can rewrite the coupling function in Eq. \eqref{Qcov} as

\begin{equation}
Q = \frac{a^{-2} \rho_c}{2C} \left[ D_{, \phi} \phi'^2  + a^2 C_{, \phi} -  2 \frac{D C_{, \phi}}{C} \phi'^2  + 2D  \left( \phi'' +  \frac{\rho_c'}{\rho_c} \phi' + 2 \mathcal{H} \phi' \right) \right],
\label{Q0dc}
\end{equation}

\noindent where we have defined the conformal Hubble rate as

\begin{equation}
    \mathcal{H} \equiv \frac{a'}{a} = a H
    \label{hubble}
\end{equation}

\noindent and $H = \dot{a}/a$ is the Hubble parameter defined in terms of the cosmic time. From the definition of the conformal and disformal functions in Eq. \eqref{transh}, Eq. \eqref{Q0dc} becomes

\begin{equation}
Q = a^{-2} h \rho_c \left[ \frac{3}{4} \frac{h_{, \phi}}{h} \phi'^2  - \frac{a^{2} h_{, \phi}}{4h^2} + \phi'  \left( 2\mathcal{H} + \frac{\rho_c'}{\rho_c} \right) +  \phi''  \right].
\label{Q0}
\end{equation}

\noindent The equation of motion for the scalar field, Eq. \eqref{eqmogen}, together with \eqref{Q0dc} yields

\begin{equation}
\phi''  -  \mathcal{H} \left( 1-3 \gamma^{-2} \right) \phi' + \frac{h_{, \phi}}{2 h^2} a^2 \left( 1- 3 \gamma^{-2} + 2\gamma^{-3} \right) + \gamma^{-3} a^2 \left( V_{, \phi} + Q \right) =0,
\label{phidd}
\end{equation}

\noindent with $Q$ as defined in Eq. \eqref{Q0}. Since the warp factor of the brane, $h(\phi)$, is always non-negative, we have $ \gamma \geq 1$. The nonrelativistic limit is realised when $a^{-2} h \phi'^2 \ll 1$ and $\gamma \rightarrow 1$. Also, it is straightforward to see that, in this limit, the scalar field Lagrangian in Eq. \eqref{action} reduces to the canonical quintessence case. Additionally, this system can also be reduced to tachyon cosmology, given an appropriate redefinition of the scalar field and the potential $V(\phi)$ \cite{Garousi:2000tr,Padmanabhan:2002cp,Copeland:2004hq}.
It is also worth noting that, even though we can state that the scalar field action in Eq. \eqref{action} is a particular case of the k-essence action \cite{PhysRevD.62.023511,ArmendarizPicon:2000ah}, the DBI scalar field cannot feature a negative pressure in the absence of the potential. From Eq. \eqref{wphi}, we easily conclude that, in the limit where the potential vanishes, we have $w_{\phi} \rightarrow 1 /\gamma$, which is always non-negative. On the other hand, when the contribution coming from the kinetic term is negligible, \textit{i.e.}, in the slow-roll limit, the scalar field exhibits a cosmological constant type of behaviour, with  $w_{\phi} \rightarrow -1$. Therefore, the coupled DBI model presents a very general framework that encapsulates some of the models studied in the literature, such as coupled quintessence \cite{vandeBruck:2015ida, vandeBruck:2017idm, Teixeira:2019hil} and coupled tachyonic dark energy \cite{Gumjudpai:2005ry,Teixeira:2019tfi}. We can immediately see that the equation of motion for the DBI scalar field is much more intricate than in the canonical case and that, taking the appropriate limit, the latter is recovered.

 From the Einstein equations, we can compute the standard Friedmann equations, characterized by the evolution of each $i$th fluid:

\begin{equation}
\mathcal{H}^2 = \frac{\kappa^2 a^2 }{3} \sum_i \rho_i = \frac{a^2 \kappa^2}{3} \left(\rho_r + \rho_b + \rho_c + \rho_\phi \right)
\label{fried1}
\end{equation}

\noindent and

\begin{equation}
\mathcal{H}' + \mathcal{H}^2 = -\frac{a^2 \kappa^2}{6} \sum_i \left(\rho_i + 3 p_i \right) = -\frac{ \kappa^2 a^2 }{6} \left(\rho_r + \rho_b + \rho_c + \rho_\phi + 3 p_r + 3 p_{\phi} \right),
\end{equation}

\noindent where we have included the other noninteracting matter components of the Universe: baryons and radiation (we assume massless neutrinos that can be incorporated together with photons in an effective radiation fluid), denoted by the subscripts $b$ and $r$, respectively. Being noninteracting, these fluids evolve according to standard conservation relations

\begin{equation}
\rho_r' + 4 \mathcal{H} \rho_r= 0,
\label{rhor}
\end{equation}

\begin{equation}
\rho_b' + 3\mathcal{H}\rho_b= 0,
\label{rhob}
\end{equation}

\noindent where we have made use of the fact that $p_b=0$ and $p_r=1/3\, \rho_r$ for each fluid. On the other hand, from Eqs. \eqref{emtcons} and \eqref{cons} we derive the continuity equations, describing the interaction between the disformally coupled fluids

\begin{equation}
\rho_{\phi}' + 3 \mathcal{H}  \rho_{\phi} \left( 1 + w_{\phi} \right) = - Q \phi',
\label{rhophi}
\end{equation}

\noindent with $w_{\phi}$ as defined in Eq. \eqref{wphi}, and 

\begin{equation}
\rho_c' + 3 \mathcal{H} \rho_c= Q \phi'.
\label{rhocdm}
\end{equation}

\noindent Note that, in the nonrelativistic limit, we have $\gamma \rightarrow 1+ \frac{h \phi'^2}{2a^2} $, and the EoS parameter for quintessence is recovered. However, in this work, we are interested in the effects coming from the noncanonical behaviour, emerging from the relativistic signatures. Combining Eqs. \eqref{Q0} and \eqref{phidd} we arrive at an expression for $Q$, containing first-order derivatives of the scalar field only: 

\begin{equation}
Q = - \left[ \frac{ h \left( V_{, \phi} + 3a^{-2} \mathcal{H} \gamma \phi' \right) + \frac{h_{, \phi}}{h} \left( 1-\frac{3}{4} \gamma \right) }{\gamma + h \rho_c} \right] \rho_c.
\label{Qnod}
\end{equation}

\noindent The sign of $Q$ in Eqs. \eqref{rhophi} and \eqref{rhocdm} determines the direction in which energy is being transferred, \textit{i.e.}, if it is the dark energy fluid that grants energy to the disformal dark matter or the other way around. It is interesting to note that the coupling has a different interpretation for each fluid. For the dark energy fluid, the coupling can be combined with the self-interacting potential to give an effective scalar field potential $V_{\rm eff} (\phi,\phi')$. On the other hand, as previously mentioned, the coupling can also be interpreted as a local change in the geometry, encoded in $\bar{g}$, defining the geodesics according to which dark matter is propagating. Instead of working with $Q$, it is convenient to define an effective coupling, $\beta$, given by

\begin{equation}
\beta \equiv \frac{Q}{\kappa \rho_c}.
\label{beta}
\end{equation}

Because cosmological constraints often assume a noninteracting dark sector, where dark matter is modeled as a noninteracting pressureless perfect fluid, we define instead an effective equation of state parameter for dark energy $w_{\phi, {\rm eff}}$ \cite{Das:2005yj}. In doing so, we map the coupled model presented here to an uncoupled framework, where dark matter is not interacting with dark energy, and all the effects of the coupling are included in an effective dark energy fluid, defined as

\begin{equation}
    \rho_{\phi, {\rm eff}} = \rho_{\phi} + \rho_c - \rho_{c,0} a^{-3}.
    \label{rhophieff}
\end{equation}

\noindent Henceforth, a subscript $0$ will be used to denote present values.

\noindent Consequently, the Friedmann equation may be written as

\begin{equation}
    \mathcal{H} = \frac{\kappa^2 a^2}{3} \left(\rho_{r,0} a^{-4} +  \rho_{b,0} a^{-3} + \rho_{c,0} a^{-3} + \rho_{\phi, {\rm eff}} \right)
\end{equation}

\noindent where $\rho_{r,0}$, $\rho_{b,0}$, and $\rho_{c,0}$, are the measured radiation, baryon, and dark matter energy densities at the present, respectively. 
%Note how the part of $\rho_c$ that doesn't evolve as pressureless matter has been included in the effective dark energy fluid, and the dark components have been effectively isolated. 
Taking the derivative of Eq. \eqref{rhophieff}, the continuity equation for the effective fluid becomes

\begin{equation}
    \rho_{\phi, {\rm eff}}' + 3 \mathcal{H} \rho_{\phi, {\rm eff}} (1+ w_{\phi, {\rm eff}}) = 0,
\end{equation}

\noindent in agreement with a standard uncoupled scenario. Comparing the previous equations with Eqs. \eqref{rhophi} and \eqref{rhocdm}, we find 

\begin{equation}
    w_{\phi, {\rm eff}} = \frac{p_{\phi}}{\rho_{\phi, {\rm eff}}},
    \label{weffphi}
\end{equation}

\noindent with $p_{\phi}$ as defined in Eq. \eqref{rhopphi}.
This definition allows for a direct comparison between the background evolution of the disformally coupled dark sector in this model and experimental data.

\subsection{Qualitative dynamics and initial conditions}

In this work, we consider the case of an AdS$_5$ throat with a quadratic potential, for which 

\begin{equation}
h(\phi) = h_0 \frac{1}{\phi^4},\ \ \ \ V(\phi) = V_0 \frac{\phi^2}{\kappa^2},
\label{hv}
\end{equation}

\noindent with $h_0, V_0 >0$. This model has four free parameters that control the cosmological evolution: the theoretical quantities associated with the scale of the warp factor and the potential, $h_0$ and $V_0$, respectively, and the initial conditions for the scalar field and its velocity, $\phi_{\rm ini}$ and $\phi'_{\rm ini}$, respectively. In Ref. \cite{Koivisto:2013fta}, it was argued that $V_0$ and $h_0$ can be combined in a single dimensionless quantity 

\begin{equation}\label{Gamma0}
  \Gamma_0 \equiv V_0\, h_0.   
\end{equation}

This is a key parameter for the dynamics of the system near the fixed point solutions, and, therefore, it is important for the late time and future cosmological evolution.
Since $C$ and $D$ are merely functions of the warp factor and tension of the brane, the disformally coupled model considered here has the same number of parameters as the corresponding standard DBI uncoupled case \cite{Guo:2008sz,Gumjudpai:2009uy,Copeland:2010jt}. However, this also means that the conformal and disformal effects cannot be directly disentangled and that we are not able to recover the uncoupled scenario by means of limit values of the parameters in the theory. 

In Ref. \cite{Koivisto:2013fta}, the background evolution of this model for different parameters was addressed through a dynamical systems analysis (see \cite{Bahamonde:2017ize} for a review on dynamical systems applied to cosmology). The main advantage is that, instead of evolving the equations numerically, a lot of information can be extracted from the fixed points of the dynamical system alone. Each specific set of initial conditions and parameters corresponds to a trajectory in the phase space, describing the evolution of the Universe. The fixed points correspond to the limit scenarios of this evolution, that describe specific periods of the Universe’s history.
One of the most appealing features related to the introduction of dark couplings is the possibility of having the emergence of scaling fixed points. These are solutions for which the two components dilute with the same rate and, hence, their fractional energy densities maintain a constant ratio, providing a more natural explanation for the observed energy share in the Universe.

From the dynamical analysis conducted in Ref. \cite{Koivisto:2013fta}, we gather that interesting cosmological settings exist when $\Gamma_0 > 1$, with the emergence of a saddle dark energy-dark matter scaling solution. A typical scenario comprises standard radiation- and matter-dominated epochs in the past followed by the scaling solution, where the system will spend a certain amount of time. During this regime, it could be the disformally coupled matter driving part of the expansion, with the accelerated expansion of the current Universe starting during a matter-dominated era, fuelled by the nonminimal coupling. In the future, and independently of the initial conditions, due to the repelling nature of the saddle point, the DDM starts diluting away and the system evolves towards a standard accelerating attractor solution, where the cosmic evolution is governed by the DBI scalar field, and all the other species become negligible.

The initial conditions for $\phi$ and $\phi'$ determine when we enter into the scaling and attractor regimes, respectively, and, the more nonrelativistic the field is, the longer the transition takes. However, these scenarios take place only in the future, since we will be considering models where the system is still approaching the scaling solution today. This means that the effects of taking different values for $\phi'_{\rm ini}$ will not be visible at the present. On the other hand, as we will see, the initial value of $\phi$ has a significant impact on the evolution of both the background and the perturbations. 
During the radiation- and matter-dominated epochs, before the transition to the scaling solution or the attractor, the system dwells in a frozen state, with $\gamma \simeq 1$ and $w_{\phi} \simeq -1$, and the DBI field resembles a cosmological constant for a broad range of initial conditions, including relativistic ones.
At late times, the system starts approaching the accelerating fixed points, characterised by $\gamma \rightarrow \infty$, at a constant rate. This rate will eventually change once the attractor is (approximately) reached. 
One interesting feature is that, for large values of $\Gamma_0$, \textit{i.e.}, for $\Gamma_0 \rightarrow \infty$, both the scaling and dark energy-dominated fixed points approach a de Sitter-like solution, resulting in a Universe totally dominated by the potential of the scalar field, with $w_{\phi} \rightarrow -1$. Therefore, we expect that, for increasingly higher values of $\Gamma_0$, the background cosmology will resemble that of a $\Lambda$CDM model near the fixed point scenarios, even though this may not be the case at the level of the perturbations.

Since the field $\phi$ represents the brane’s position in the extradimensional space, we take $\phi \geq 0$ for consistency. Furthermore, in order to be moving down the throat with $\phi \geq 0$, we take $\phi' <0$, which is also needed in order to reproduce cosmologies that allow for the scaling regime. For these reasons, we restrict this study to cases with $\{ \phi_{\rm ini}>0,\phi'_{\rm ini}<0 \}$.

In what follows, we wish to present a numerical study, highlighting some of the cosmological features in this model. We will also be interested in analysing cosmologically plausible scenarios, where the dark energy field cannot have had a significant impact in the past \cite{Pettorino:2013ia} and, therefore must resemble a slowly evolving cosmological constant. We have seen that, in this scenario, the accelerated expansion can be accounted for, in part, by the disformal dark matter, and, therefore, we expect that different signatures will be present (when compared to $\Lambda$CDM and coupled quintessence models).

\subsection{Implementation and Background Results}

\begin{table*}[t]
\parbox{.49\linewidth}{
\centering
\setlength{\tabcolsep}{2pt}
\begin{tabular}{c c c c c c c}
\hline\hline \\[-1.5ex]
Model & $\Gamma_0$ & $\phi_{\rm ini}\, ({\rm M_{pl}})$ & $\gamma^0$ & $w_{\phi}^0$ & $G_{\rm eff}^0/G_N$ & $\sigma_8$ \\ [0.5ex]% inserts table %heading
\hline \\[-1.5ex]
M$1$ & 1.5 & 3 & 1.029 & -0.799 & 1.286 & 1.014 \\
M$2$ & 5 & 3 & 1.051 & -0.870 & 1.147 & 0.892 \\ 
M$3$ & 10 & 3 & 1.063 & -0.911 & 1.038 & 0.853 \\[1ex]
\hline
\end{tabular}
}
\hfill
\parbox{.49\linewidth}{
\centering
\setlength{\tabcolsep}{2pt}
\begin{tabular}{c c c c c c c}
\hline\hline \\[-1.5ex]
Model & $\Gamma_0$ & $\phi_{\rm ini}\, ({\rm M_{pl}})$ & $\gamma^0$ & $w_{\phi}^0$ & $G_{\rm eff}^0/G_N$ & $\sigma_8$ \\ [0.5ex]% inserts table %heading
\hline \\[-1.5ex]
M$4$ & 1.5 & 1.7 & 1.434 & -0.601 & 1.004 & 0.916 \\
M$5$ & 5 & 1.7 & 1.306 & -0.819 & 1.172 & 0.797 \\ 
M$6$ & 10 & 1.7 & 1.254 & -0.904 & 1.976 & 0.804 \\[1ex]
\hline
\end{tabular}
}
\caption{\label{tabel1} Parameter values for the illustrative models considered in this work. The initial condition for the velocity of the scalar field is $\phi'_{\rm ini} = - 10^{-25} ({\rm Mpl\, Mpc}^{-1}) $ for all the models considered. In all cases, we assume a spatially flat Universe and Planck cosmological reference parameters \cite{Aghanim:2018eyx} have been chosen for the present time: $\Omega_{b,0} h^2 = 0.022032$, $\Omega_{c,0} h^2 = 0.12038$, $T_{CMB}=2.7255$ K, and $H_0 = 67.556$ km/s/Mpc. 
The models considered in this work are illustrative only and not necessarily cosmologically  viable. We also present some relevant quantities, where a superscript $0$ denotes present values. The value of the Lorentz factor, $\gamma^0$, quantifies present deviations from standard quintessence. The value of the DBI EoS parameter, $w_{\phi}^0$, is a measure of present deviations from cosmological-constant-like behaviour. We also show the enhancement in the effective gravitational coupling $G_{\rm eff}^0/G_N$ (defined in Eq. \eqref{geff}) and the value of the predicted $\sigma_8$ value for each model.}
\end{table*}

We have implemented the model in a modified version of the Boltzmann code CLASS \cite{Lesgourgues:2011re, Blas:2011rf}. This allows for a complete study of the cosmological predictions of the theory, and, from there, it is possible to infer what is the interesting range of parameter values to consider. We adapted the code to account for the background and the linear perturbations for a DBI scalar field. For the simulations we adopt the Planck 2018 cosmological parameters \cite{Aghanim:2018eyx}: $\Omega_{b,0} h^2 = 0.022032$, $\Omega_{c,0} h^2 = 0.12038$, $T_{\rm CMB}=2.7255$ K, and $H_0 = 67.556$ km/s/Mpc, on a spatially flat cosmological background. 
The scale of the potential, $V_0$, is taken to be a shooting parameter that is numerically adjusted such that we recover the fiducial Planck cosmological value, $\Omega_{\phi} \simeq 0.68$, for the fractional energy density of the scalar field, while assuming that we are very close to the scaling regime today. 

\begin{figure}
      \subfloat{\includegraphics[height=0.295\linewidth]{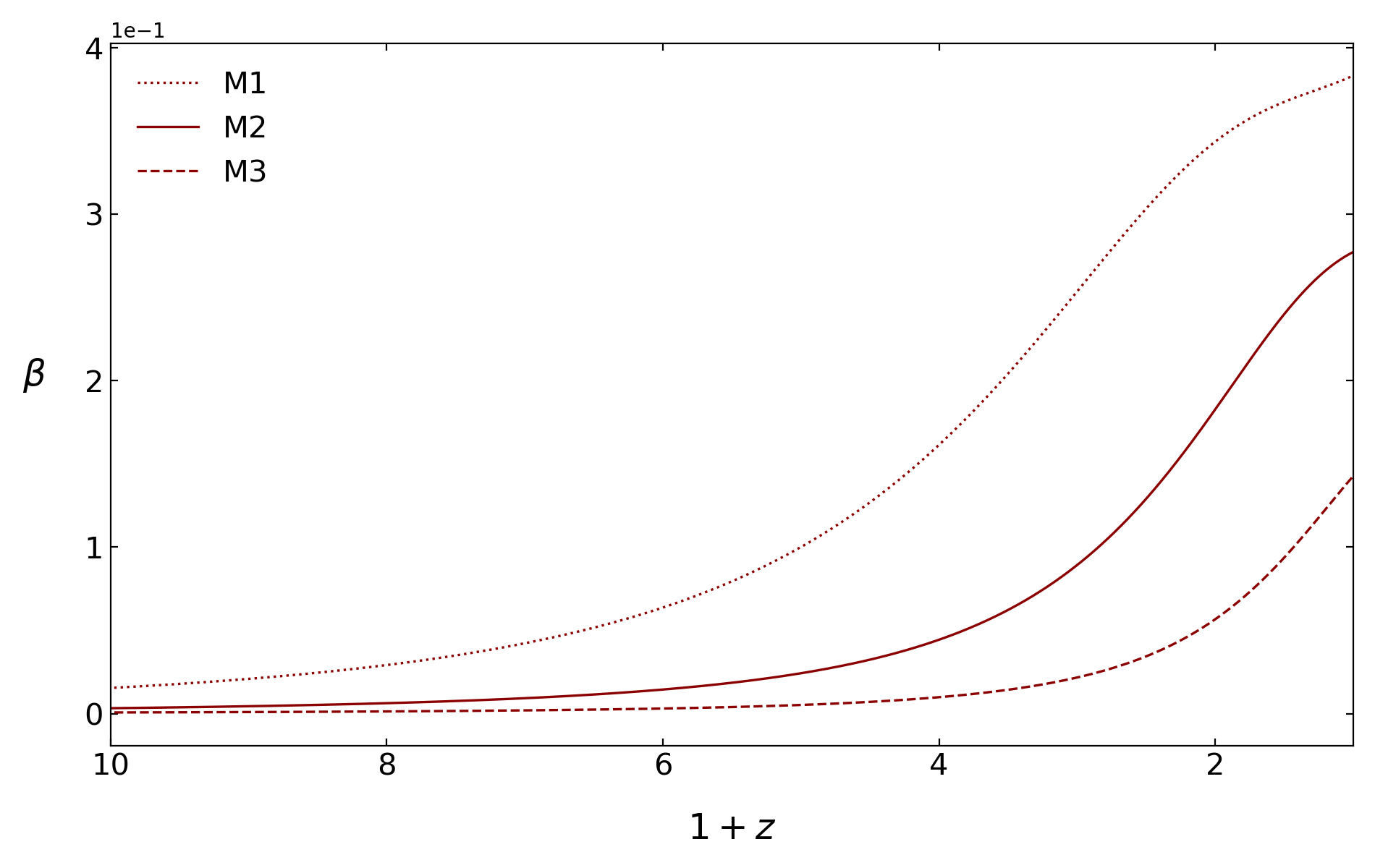}}
      \hfill
      \subfloat{\includegraphics[height=0.295\linewidth]{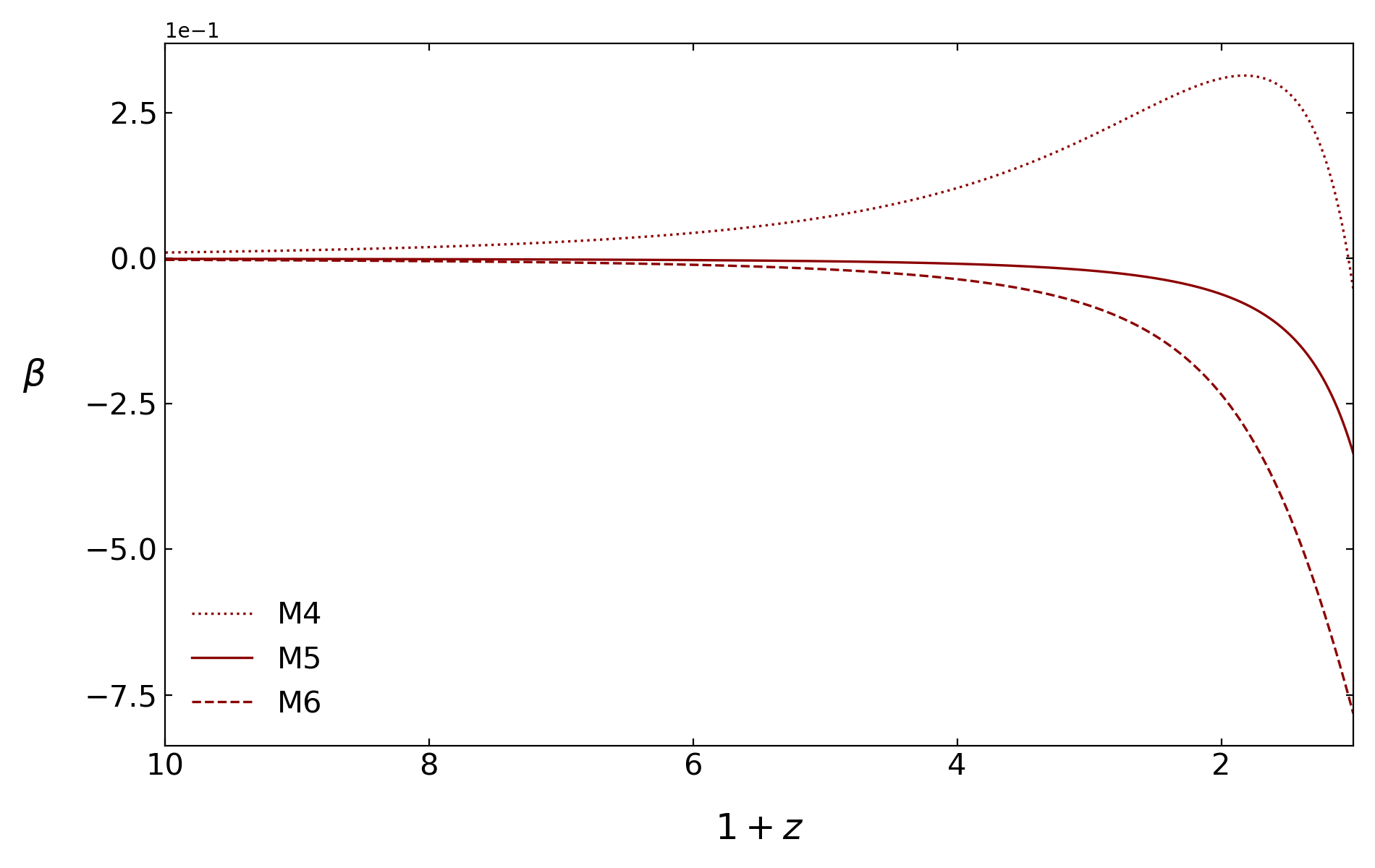}} 
  \caption{\label{fig:difgamma0beta} Background evolution of the effective coupling function $\beta$, defined in Eq. \eqref{beta}. The left and right panels correspond to Models M1-M3 and M4-M6 in Table \ref{tabel1}, respectively. All quantities are plotted as functions of the redshift $z$, related to scale factor as $1+z = a_0/a$. We clearly identify two regimes of the theory: one in which the coupling is always positive throughout the cosmic evolution, depicted on the left panel, and another one for which the coupling may start out as being positive but eventually starts to decrease towards negative values at the present, pictured on the right panel.}
\end{figure}

We studied different regimes of the model, characterised by different initial conditions and different values of $\Gamma_0$. The parameter combinations considered for illustration purposes are presented in Table \ref{tabel1}. 
We consider two groups of coupled models, characterised by different initial values of the scalar field, that give rise to distinct natures of the effective coupling, as depicted in Fig. \ref{fig:difgamma0beta}, for all the models in consideration. The first group is represented here by the case study in the left-hand side table in Table \ref{tabel1}, in which $\phi_{\rm ini} = 3 {\rm M_{pl}}$. In this case the effective coupling is always positive up until the present and grows more rapidly for smaller values of $\Gamma_0$. On the other hand, for the second group, exemplified in the right-hand side table of Table \ref{tabel1}, with $\phi_{\rm ini} = 1.7 {\rm M_{pl}}$, the coupling is always negative today but may start out as being positive, for a significant part of the evolution, as is the case for the model M4 in Table \ref{tabel1}. In contrast, for the second group, the value of the coupling becomes greater (in absolute value) for higher values of $\Gamma_0$.

\begin{figure}
       \subfloat{\includegraphics[height=0.29\linewidth]{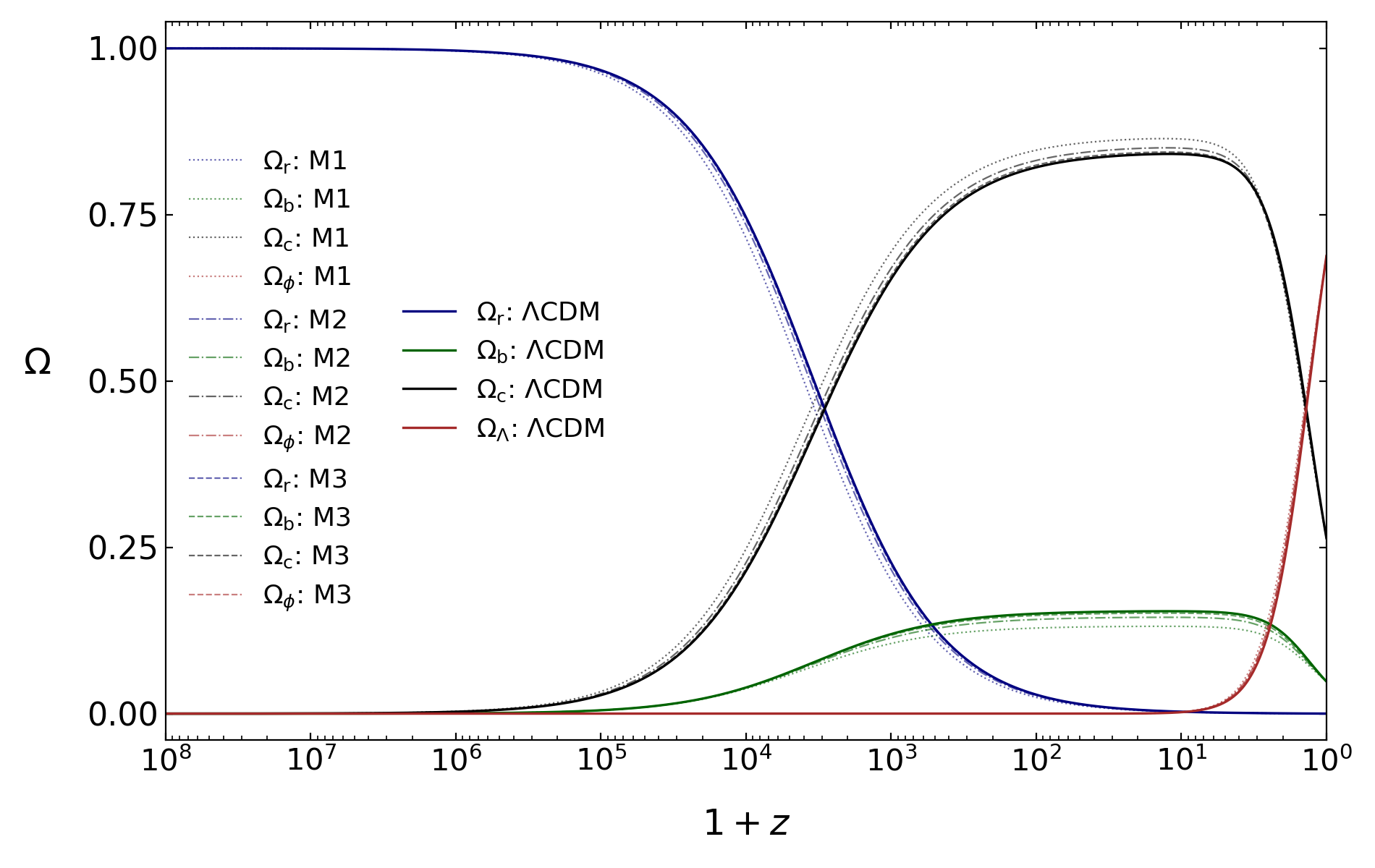}} 
       \hfill
      \subfloat{\includegraphics[height=0.29\linewidth]{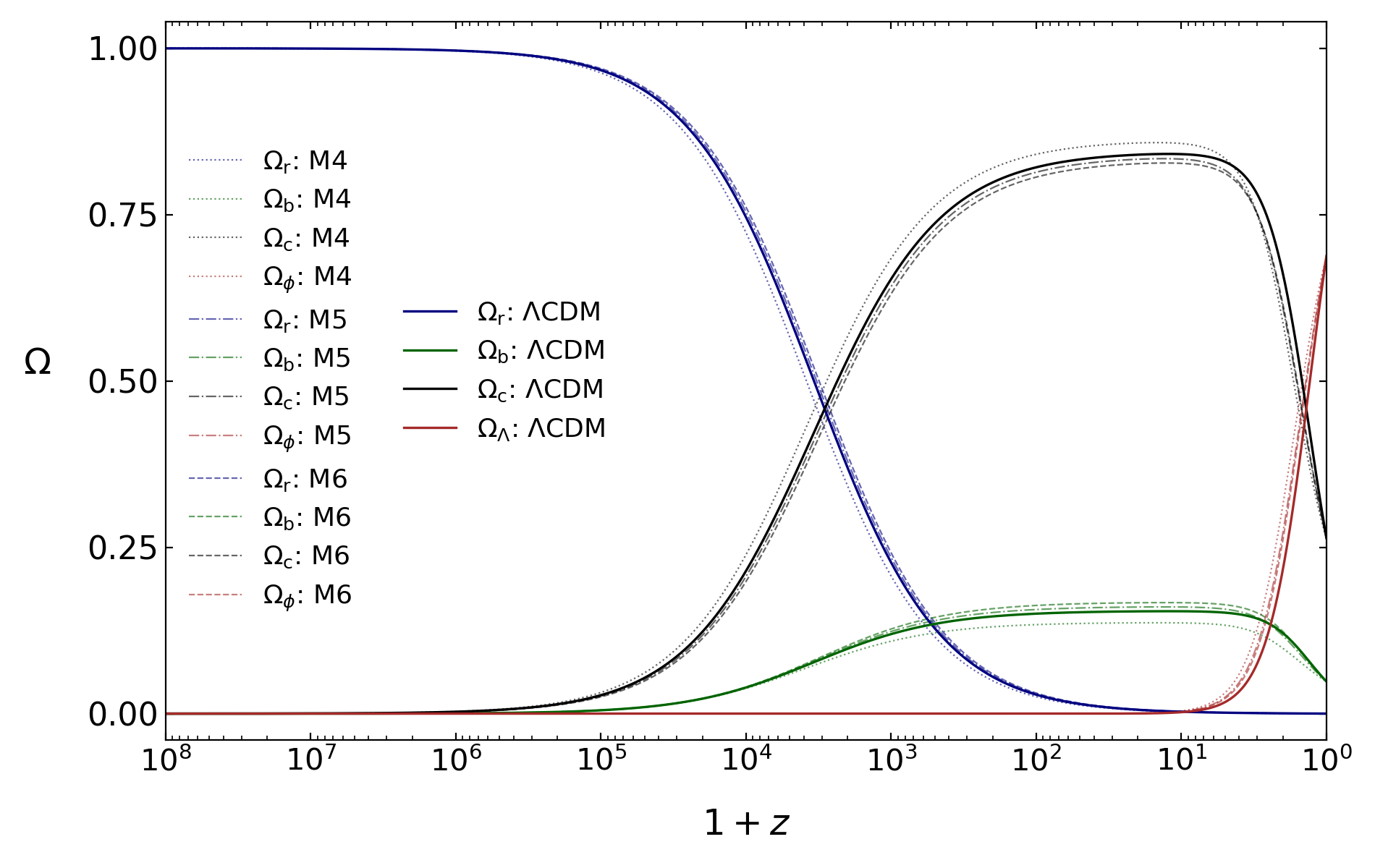}}
             \hfill
       \subfloat{\includegraphics[height=0.29\linewidth]{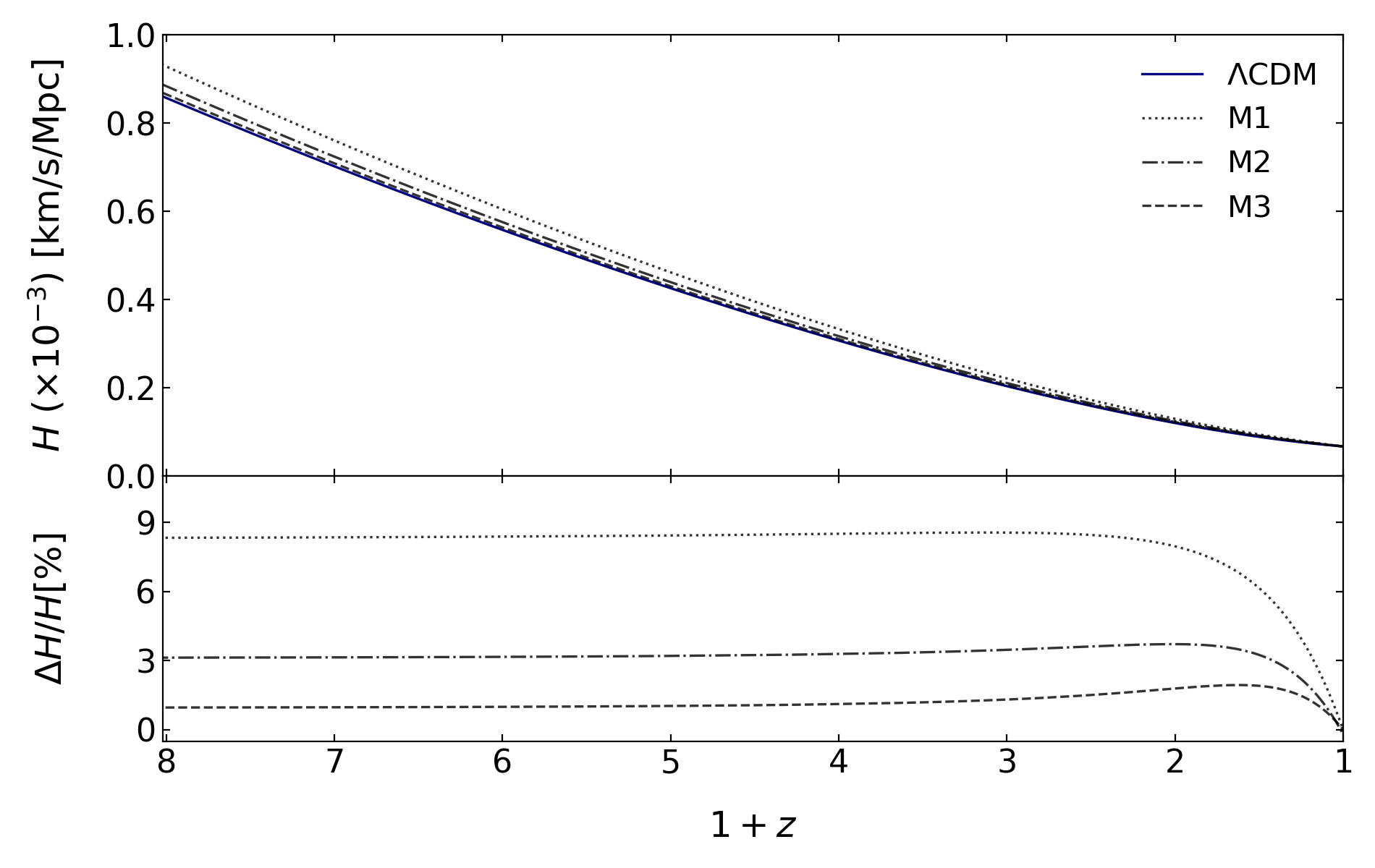}}
                          \hfill
             \subfloat{\includegraphics[height=0.29\linewidth]{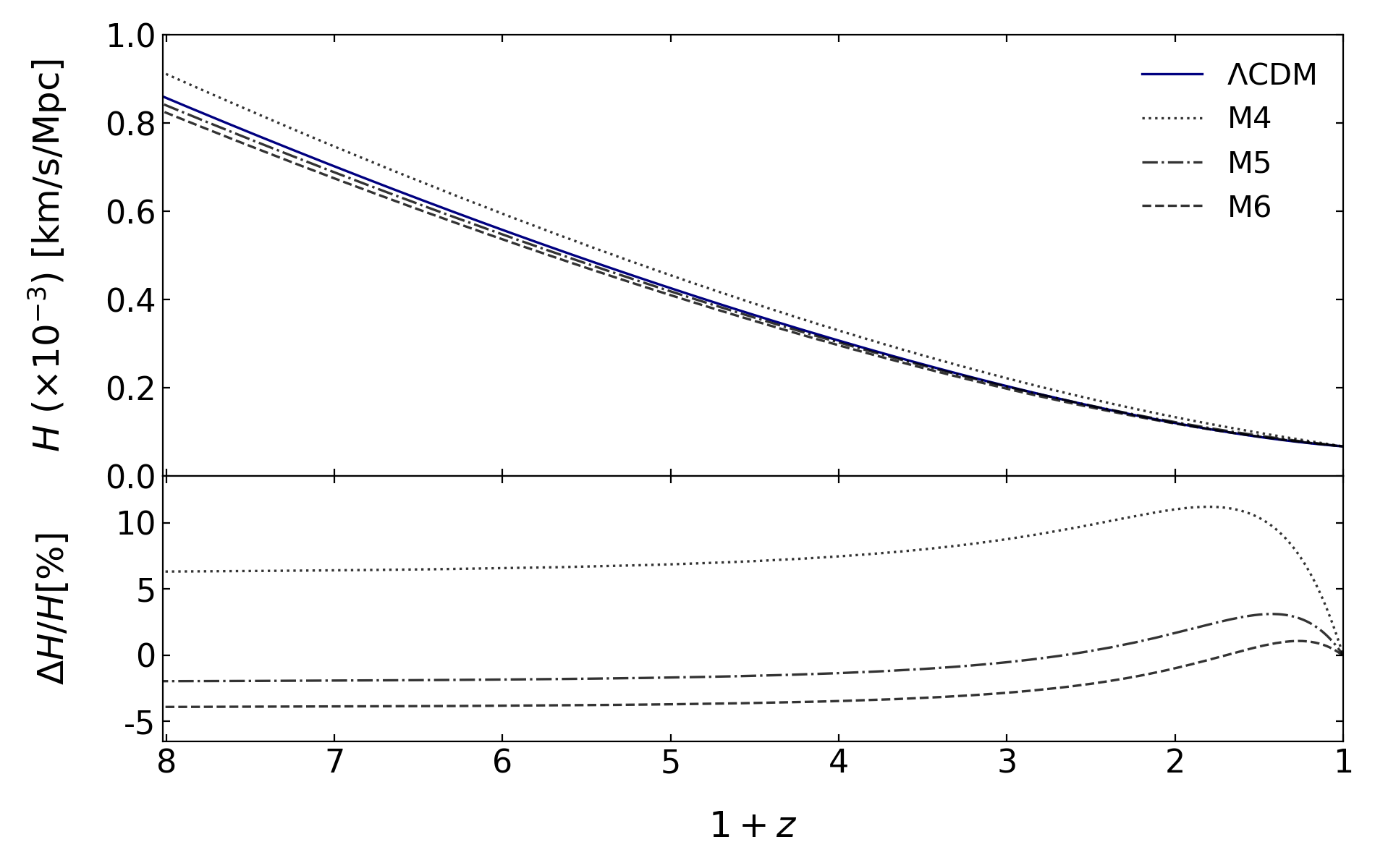}}
                       \hfill
      \subfloat{\includegraphics[height=0.29\linewidth]{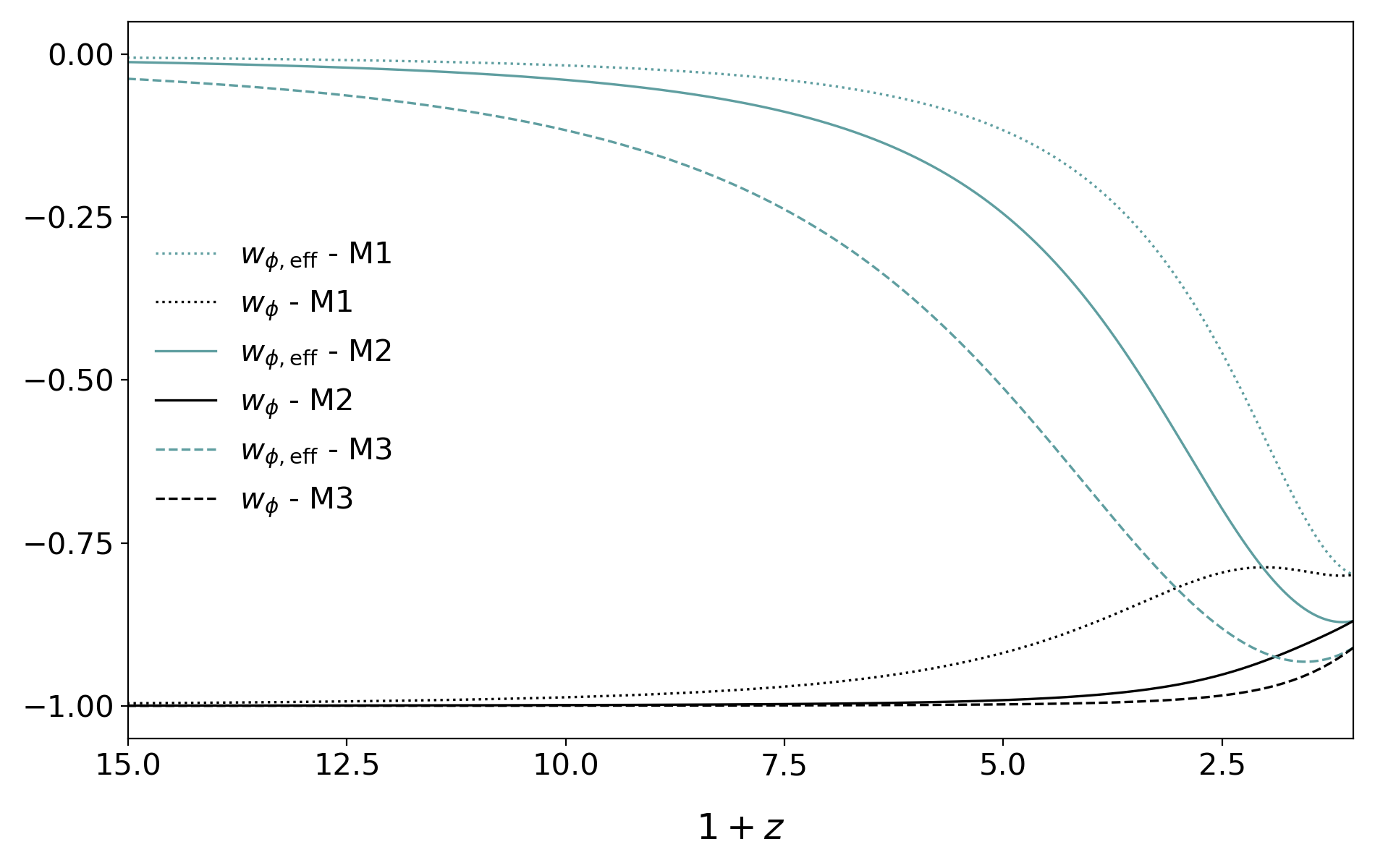}}
                         \hfill
            \subfloat{\includegraphics[height=0.29\linewidth]{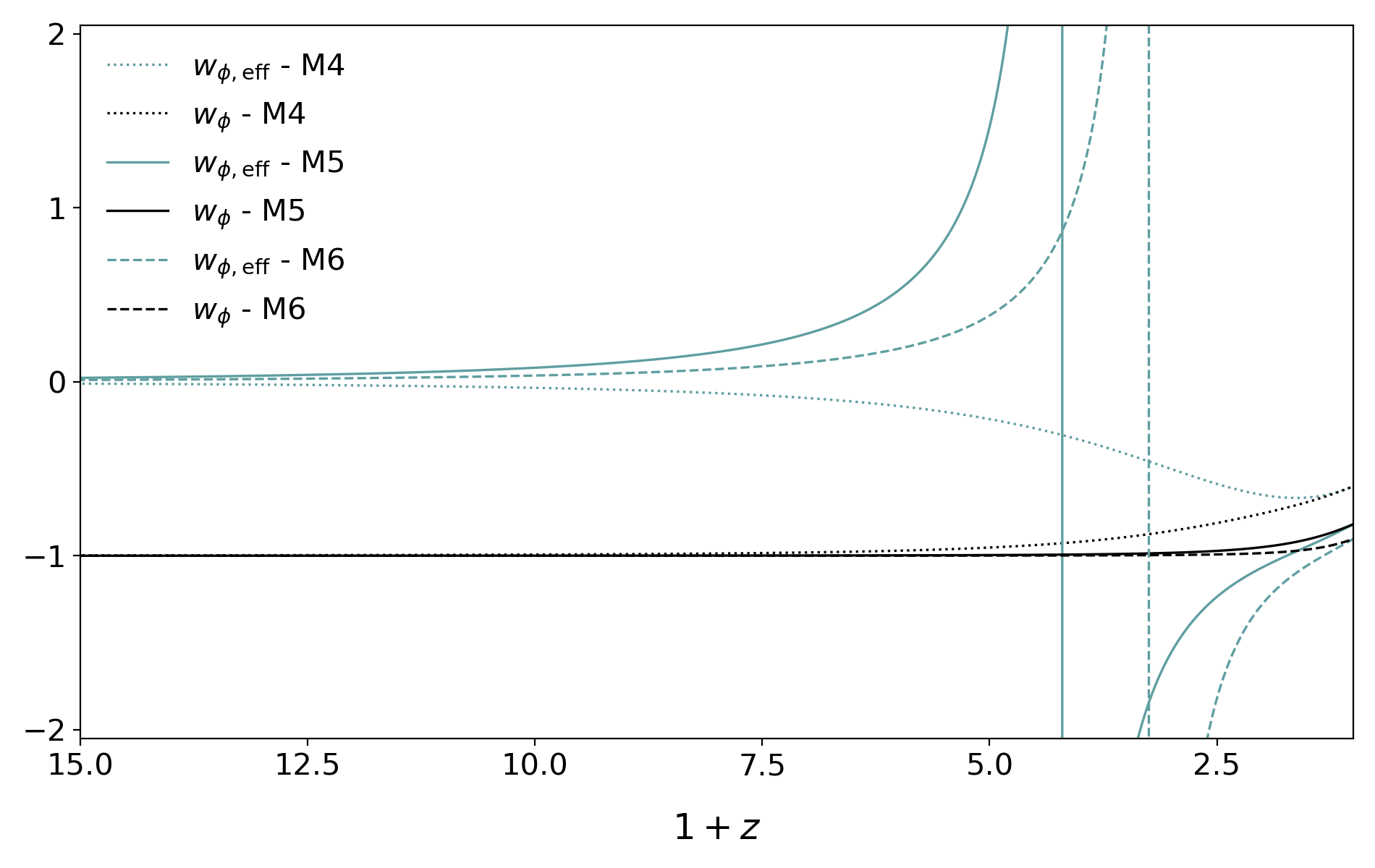}}
  \caption{\label{fig:difgamma0} Background evolution of the fractional energy densities $\Omega_i = k^2 \rho_i/3 H^2$, for each labelled $i$th fluid (top), of the Hubble expansion rate $H$ (middle), defined in Eq. \eqref{hubble}, and the DBI and effective equation of state parameters, as defined in Eqs. \eqref{wphi} and \eqref{weffphi} (bottom), as functions of the redshift, $z$. In the middle panel, we also present the deviations of the Hubble expansion rate in each model, $\Delta H$, when compared to the concordance model. The left and right panels correspond to models M1-M3 and M4-M6 in Table \ref{tabel1}, respectively.}
\end{figure}

We find that there is no simple degeneracy between $\Gamma_0$ and $\phi_{\rm ini}$. Since  the warp factor has an inverse dependence on the value of the scalar field, for higher values of $\phi_{\rm ini}$, $h_0$ has to be higher (and, consequently, $V_0$ smaller), given the same $\Gamma_0$ value, in order to obtain the fiducial cosmology at the present. 

We now wish to study the redshift evolution of some relevant background quantities for the cosmological models considered in this work: the effective coupling $\beta$, as defined in Eq. \eqref{beta}; the fractional energy densities $\Omega_i$ for each $i$th fluid, defined as $\Omega_i = \kappa^2 \rho_i/3 H^2$; the Hubble expansion rate $H$, defined in Eq. \eqref{hubble}; and the DBI and effective equation of state parameters, as defined in Eqs. \eqref{wphi} and \eqref{weffphi}. These are standard quantities that characterise the background evolution of the Universe. Moreover, these will also be relevant for the study of the linear growth rate of structures, strongly entangled with the background evolution.

The aforementioned background quantities are depicted in Figs. \ref{fig:difgamma0beta} and \ref{fig:difgamma0}. We refer to the captions and to the figures themselves for the relevant details and for the different colours and/or line styles used. In the left panels, we depict the models M1-M3 in Table \ref{tabel1}, for which the effective coupling is always positive. However, we expect that, as we approach the scaling solution, characterised by $1/\gamma \rightarrow 0$, the effective coupling should become negative. On the other hand, in the right panels, we plot the models M4-M6 in Table \ref{tabel1}, in which $\beta$ is always negative at late times. It is worth noting that model M4 has a very characteristic evolution for the effective coupling, when compared to M5 and M6: $\beta$ starts growing much earlier on towards positive values, reaching a peak value around $z = 0.8 $ ($\beta \approx 0.3$), and then starts decreasing, becoming negative and significantly smaller when $z=0$ ($\beta \approx -0.05$). This shows that, in principle, this model allows for scenarios in which, although the coupling in the dark sector is negligible at the present, it may have been significant over the past evolution.
Our simulations suggest that, for the highest value of $\phi_{\rm ini}$ (right panel), the coupling is activated earlier, when compared to the same values of $\Gamma_0$. We also observe that, when $\beta>0$ ($<0$), for higher values of $\Gamma_0$ the coupling is activated at smaller (higher) redshifts, leading to smaller (higher) absolute values for the coupling today. This is also consistent with cosmologies growing closer to (away from) $\Lambda$CDM during the matter dominated epoch, which clearly illustrates the differences between the two regimes. However, we expect that, once we start approaching the scaling solution, for the models M1-M3, the coupling will become negative independently of the initial conditions, with a characteristic turning point, as the one observed in M4.

In the top panels of Fig. \ref{fig:difgamma0} we show the evolution of the fractional energy densities. The past evolution of the Universe is as explained in the previous section: a radiation-dominated epoch, followed by a DDM-dominated epoch, with the dark energy fluid becoming important at the present, as the system starts evolving towards the scaling solution. As an artefact of the shooting method performed in CLASS, we find that the matter-radiation equality is also shifted, despite the fact that the radiation fluid is uncoupled and both the coupling and the DBI energy density are negligible at this epoch. At late times, we find that there is effectively a deviation of the fractional energy densities from $\Lambda$CDM, with higher (lower) abundances of dark matter for positive (negative) couplings, accounting for the transfer of energy between the dark fluids. As for the evolution of $\Omega_{\phi}$, we clearly see that the dashed curves (corresponding to the highest value of $\Gamma_0$ considered) are the closest to $\Omega_{\Lambda}$ in both regimes. 

The present value of the expansion rate is fixed according to the Planck fiducial value  ($H_0 = 67.556$ km/s/Mpc). However, the $H(z)$ functions are different in all of the cases considered, and due to this effect, different models predict different evolutions for the fractional energy densities of the uncoupled species, that is, baryons and radiation. In the middle panels of Fig. \ref{fig:difgamma0} we see that negative (positive) couplings are associated with suppressed (enhanced) expansion rates, that approach the $\Lambda$CDM one for lower (higher) values of $\Gamma_0$. The evolution of $\Delta H(z)$ for each case is also presented in the bottom of the middle panels. Interestingly, close to the present ($z\lesssim 0.5$), we find $\Delta H(z) > 0$ for all models, meaning that the Hubble rate is always larger than the $\Lambda$CDM reference values. For models M5 and M6, at higher redshifts, $H(z)$ is smaller than the concordance model. For the scenarios with positive (negative) couplings, in the cases with higher (lower) $\Gamma_0$, the Hubble rate approaches the $\Lambda$CDM curve for higher redshifts, to the point where the differences become very small. This result is interesting, because it means that, in principle, by taking increasingly higher (lower) values of $\Gamma_0$, models with positive (negative) couplings may approach $\Lambda$CDM with increasingly higher accuracy, even if the coupling is significantly higher during the past or present evolution (such as in M4). Even though the models chosen present high couplings, when compared with present constraints, we see that the differences when compared to the concordance model are always below $10 \%$. By having a different expansion history, these models are expected to have specific observational imprints. This will be studied in detail in the next section, where we will see that variations in the evolution of the Hubble parameter, together with the effect of changing the value of the effective gravitational constant, produces particular signatures in the linear growth of structures.

Also, in compliance with the study of the effective equation of state for dark energy (bottom panels of Fig. \ref{fig:difgamma0}) conducted in the previous section, we find that for positive values of $\beta$ phantom behaviour, $w_{\phi}<-1$, is never observed during the past evolution of the scalar field (models M1-M4). On the other hand, for negative values of $\beta$, we find that the phantom behaviour does emerge, with the transition taking place earlier for smaller values of $\Gamma_0$ (models M5 and M6). Higher values of $\Gamma_0$ also lead to values of $w_{\phi}^0$ increasingly closer to $-1$, consistent with a cosmological constant in the limit $\Gamma_0 \rightarrow \infty$. On the other hand, for the models where the coupling turns on earlier (M1 and M4 in each regime), $w_{\phi}$ also starts departing from $-1$ at earlier redshifts.

So far, everything seems to be as expected from the dynamical and qualitative studies, but we expect that distinguished features should arise at the linear level of cosmological perturbation theory, analysed in the following section, under the regimes specified above.

\section{Perturbations} \label{sec:pert}

In this section, we will discuss the evolution of cosmological perturbations. From the background analysis, we expect that the dark D-brane model will feature a rich phenomenology at the linear level, allowing one to probe predictions of the theory that may be constrained with the available observational data. We calculate the spectrum of anisotropies in the CMB and the predictions for the matter power spectrum. The equations in this section are given in the Newtonian gauge, specifically for the model discussed in this paper. For completeness, the expressions in the synchronous gauge are given in Appendix \ref{app:synch}, along with the equations for generic conformal coupling $C(\phi)$ and general disformal coupling $D(\phi)$. 

We focus on scalar perturbations in the conformal Newtonian gauge, in which the perturbed line element is given by 

\begin{equation}
ds^2= a^2(\tau) \left[ - \left( 1 + 2 \Psi \right) d\tau^2 + \left( 1-2\Phi\right) \delta_{ij} dx^{i} dx^{j} \right].
\label{pertmet}
\end{equation}

\noindent The quantities $\Psi (\tau, x^{i})$ and $\Phi (\tau, x^{i})$ are the scalar metric perturbations, and $\delta_{ij}$ is the Kronecker delta symbol, with Latin indices denoting spatial coordinates. 
From Eqs. \eqref{eins} and \eqref{pertmet}, it is straightforward to compute the components of the perturbed Einstein equations:
\begin{equation}
\delta G^{\mu}_{~\nu} = \kappa^2 \delta T^{\mu}_{~\nu},
\label{perteinst}
\end{equation}
\noindent where $\delta G^{\mu}_{~\nu}$ and $\delta T^{\mu}_{~\nu}$ are the perturbed Einstein and energy-momentum tensor, respectively. For each fluid, the individual components of $\delta T^{\mu}_{~\nu}$ read

\begin{align}
    \delta T^{0}_{~0,f} &= - \delta \rho_f, \label{dt1} \\
    \delta T^{0}_{~i,f} &= (\rho_f + p_f) \partial_i v_f ,\label{dt2} \\
    \delta T^{i}_{~0,f} &= - (\rho_f + p_f) \partial^{i} v_f, \label{dt3} \\
    \delta T^{i}_{~j,f} &= \delta p_f \delta^{i}_{j} + \Pi^{i}_{~j,f}, \label{dt4}
\end{align}

\noindent where $f$ is an index for each individual fluid and $\delta \rho_f$, $\delta p_f$, $v_f$, and $\Pi^{i}_{j,f}$ stand for the perturbation of the energy density, the perturbation of the pressure, the peculiar velocity potential, and the anisotropic stress tensor of the fluid $f$, respectively. If the matter source is specified, then the perturbation of the energy-momentum tensor is defined accordingly. For the model in consideration, the perturbed Einstein equations, written in Fourier space, read

\begin{align}
k^2 \Phi + 3 \mathcal{H} \left( \Phi'+ \mathcal{H} \Psi \right) &= -4 \pi G_N a^2 \sum_f \delta \rho_f,
\label{peq1} \\
k^2 \left( \Phi' + \mathcal{H} \Psi \right) &= 4 \pi G_N a^2 \sum_f \rho_f \left( 1 + w_f \right) \theta_f,
\label{peq2} \\
\Phi'' + \mathcal{H} \left( \Psi' + 2 \Phi' \right) + \Psi \left( \mathcal{H}^2 + 2 \mathcal{H}' \right) + \frac{k^2}{3} \left( \Phi - \Psi \right) &= 4 \pi G_N a^2 \sum_f \delta p_f,
\label{peq3} \\
k^2 \left( \Phi - \Psi \right) &= 12 \pi G_N a^2 \sum_f \rho_f \left( 1 + w_f \right) \sigma_f~.
\label{peq4}
\end{align}

\noindent These equations relate the scalar potentials $\Phi$ and $\Psi$ to the perturbations in the matter fluids. The metric potentials have been expanded in a Fourier transform, which in practice translates to the replacement of spatial derivatives by the Fourier mode for each wave number, $k$. In the equations above, we have also defined the velocity potential, $\theta_f= \partial^i \partial_i v_f$, and a re-scaled anisotropic stress perturbation, $\sigma_f = \frac{2 w_f \Pi_f}{3 \left( 1 + w_f \right)}$. As a first approximation, we will consider the case of vanishing $\sigma_f$ for all fluids. According to the fourth perturbed Einstein equation, Eq. \eqref{peq4}, this implies the equality of the two gravitational potentials: $\Psi = \Phi$. In what follows, we will also replace the perturbed energy density $\delta \rho_f$ by the density contrast, defined as the perturbed energy density weighted over its background counterpart:

\begin{equation}
 \delta_f = \delta \rho_f / \rho_f.   
 \label{deltan}
\end{equation}
We also consider that all the fluids are characterised by an adiabatic speed of sound:
\begin{equation}
c_{s,f}^2 = \delta p_f / \delta \rho_f.
\label{csn}
\end{equation}
The perturbed conservation equations are derived from the energy conservation equation, $\nabla_{\nu} T^{\mu\nu}_f= 0$:

\begin{equation}
    \nabla_{\mu} \delta T^{\mu}_{\nu,u} + \delta \Gamma^{\mu}_{\mu \beta} T^{\beta}_{\nu, u} -  \delta \Gamma^{\beta}_{\mu \nu} T^{\mu}_{\beta, u} = 0, \ \ \text{and}\ \ \nabla_{\mu} \delta T^{\mu}_{\nu,c} + \delta \Gamma^{\mu}_{\mu \beta} T^{\beta}_{\nu, c} -  \delta \Gamma^{\beta}_{\mu \nu} T^{\mu}_{\beta, c} = - Q \partial_{\nu} \phi,
    \label{contpert}
\end{equation}

\noindent where the indices $f=\{u,c\}$ stand for uncoupled and coupled components, respectively, with respect to the scalar field. The quantities $\delta \Gamma^{\mu}_{\nu \beta}$ are the perturbation of the Christoffel symbols. Eqs. \eqref{contpert} can be computed using Eqs. \eqref{dt1}-\eqref{dt4}, \eqref{deltan}, and \eqref{csn} and give rise to evolution equations for the density contrast, $\delta_f$ and velocity potential, $\theta_f$, of each fluid. The baryonic and radiation sectors remain uncoupled from the scalar field and satisfy the following conservation equations, presented here for a generic barotropic fluid with equation of state $w_u =p_u/\rho_u$:

\begin{equation}
\delta_u'+ 3 \mathcal{H} \left( c_{s,u}^2-w_u \right) \delta_u = \left( 1 + w_u \right) \left( 3 \Phi' - \theta_u \right)
\label{conseu}
\end{equation}

\noindent and

\begin{equation}
\theta_u' + \left[ \mathcal{H} \left( 1 - 3 w_u \right) + \frac{w_u'}{1+w_u} \right] \theta_u = k^2 \left[ \frac{c_{s,u}^2}{1+w_u} \delta_u + \Psi \right] - k^2 \sigma_u,
\label{conseu2}
\end{equation}

\noindent with $u=\{ b,r \}$. The first equation is called the perturbed continuity equation, and the second one is the Euler equation, stemming from the time and spatial components of the energy conservation equation, respectively. 
Since DDM is the only component coupled to the scalar field, the continuity equation and the Euler equation are modified. Setting $w_c=0=c_{s,c}$, we find 

\begin{equation}\label{CDM1}
\delta_c'  = -  \left( \theta_c - 3 \Phi' \right) - \frac{Q}{\rho_c} \phi' \delta_c + \frac{Q}{\rho_c} \delta \phi' + \frac{\delta Q}{\rho_c} \phi'
\end{equation}

\noindent and

\begin{equation}\label{CDM2}
\theta_c' + \mathcal{H}  \theta_c = k^2 \Psi - \frac{Q \phi'}{\rho_c} \theta_c + k^2 \frac{Q}{\rho_c} \delta \phi,
\end{equation}

\noindent that have clear dependencies on the coupling function $Q$ and its perturbation, $\delta Q$. The evolution of the perturbed DBI scalar field is dictated by the perturbed Klein-Gordon equation:

\begin{align}\label{PerturbKG}
&\delta \phi'' + \left[ \frac{3h_{,\phi}}{h} \left( 1- \gamma^{-1} \right) \phi'    - \mathcal{H} \left( 7 - 9 \gamma^{-2}  \right)   - 3 h  \left( V_{,\phi} + Q \right) \gamma^{-1} \phi' \right] \delta \phi' + \left[-\frac{3h_{,\phi}}{h} \mathcal{H} \left( 1 - \gamma^{-2} \right) \phi' + a^2 V_{,\phi \phi} \gamma^{-3} \right. \nonumber \\
& \left. + \frac{h_{,\phi \phi}}{2h^2} a^2 \left( 1 - 3\gamma^{-2} + 2\gamma^{-3} \right) - \frac{3}{2} \frac{h_{,\phi}}{h} a^2 \left( V_{,\phi} + Q \right) \left( \gamma^{-1} - \gamma^{-3} \right) + \frac{h_{,\phi}^2}{2 h^3} a^2 \left( 1 - 3\gamma^{-1} + 3\gamma^{-2} - \gamma^{-3} \right) \right] \delta \phi  \nonumber \\
&+ \left[ 6 \mathcal{H} \left( 1- \gamma^{-2} \right) \phi'  - \frac{h_{,\phi}}{h^2} a^2  \left( 2 - 3 \gamma^{-1} + \gamma^{-3} \right) + a^2  \left( V_{,\phi} + Q \right) \left(3\gamma^{-1} -\gamma^{-3} \right) \right] \Psi - \phi'  \Psi' - 3 \gamma^{-2} \phi'  \Phi' \\
& - \gamma^{-2} \partial^i \partial_i \delta \phi+ a^2 \gamma^{-3} \delta Q=0. \nonumber
\end{align}

\noindent To derive this equation, we made use of the background Klein--Gordon equation. Finally, the perturbation of the coupling $Q$ is given in the Newtonian gauge by\footnote{The expression for general conformal, $C(\phi)$, and disformal, $D(\phi)$ coupling functions is given in Appendix \ref{app:newt}.} 

\begin{equation}
\delta Q = \frac{a^{-2} \rho_c}{\gamma^{-2} + h \rho_c \gamma^{-3} } \left( \mathcal{Q}_1 \delta_c + \mathcal{Q}_2 \Phi' + \mathcal{Q}_3 \Psi + \mathcal{Q}_4 \delta \phi' + \mathcal{Q}_5 \delta \phi \right),
\label{deltaQ}
\end{equation}

\noindent where we have defined the coefficients

\begin{flalign}
\mathcal{Q}_1 = & a^2 \frac{Q}{\rho_c}\gamma^{-2} + 3 h \frac{\delta p_c}{\delta \rho_c} \left(  a^2 \frac{h_{,\phi}}{4h^2} - \mathcal{H} \phi' \right), &
\end{flalign}

\begin{flalign}
\mathcal{Q}_2 = & 3 h \left( \gamma^{-2} + w \right) \phi', &
\end{flalign}

% \begin{flalign}
% \mathcal{B}_3 = & - 6 h \mathcal{H} \left( 1 - 2 \gamma^{-2} - w \right) \phi' - 3 a^2 h \left( V_{,\phi} + Q \right) \left( \gamma^{-1} - \gamma^{-3} \right) - 2 a^2 \frac{Q}{\rho_c} \left( 1 - \gamma^{-2} \right) \\
% &+ a^2 \frac{3}{2} \frac{h_{,\phi}}{h} \left( 1 - 2 \gamma^{-1} - \gamma^{-2} + 2 \gamma^{-3} \right), \nonumber &
% \end{flalign}

\begin{flalign}
\mathcal{Q}_3 = & 3 h \mathcal{H} \left( 1 +  \gamma^{-2} + 2w \right) \phi'  - a^2 \frac{3}{4} \frac{h_{,\phi}}{h} \left( 1 - \gamma^{-2} \right) + a^2 \frac{Q}{\rho_c} \left( 1 - \gamma^{-2} \right), &
\end{flalign}

\begin{flalign}
\mathcal{Q}_4 = & 3 h \mathcal{H} \left( 2 - 3 \gamma^{-2} - w \right) + 3 h^2 \left( V_{,\phi} + Q \right) \gamma^{-1} \phi' + 2 h \frac{Q}{\rho_c} \phi' - \frac{3}{2} h_{,\phi} \left( 1 - 2 \gamma^{-1} \right) \phi', &
\end{flalign}

\begin{flalign}
\mathcal{Q}_5 = & -k^2 h \left( \gamma^{-2} + w \right) + a^2 \frac{h_{,\phi}^2}{2 h^2} \left( \frac{3}{4} -\frac{15}{4} \gamma^{-2}  + 4 \gamma^{-3} - \frac{3}{2} w \right) + a^2 \frac{3}{4} \frac{h_{,\phi \phi}}{h} \left( \gamma^{-2} - \frac{4}{3} \gamma^{-3} + w \right)  \label{q5} \\
& - \frac{3}{2} h_{,\phi} \mathcal{H} \left( 1 - \gamma^{-2} +2w \right) \phi' - a^2 h V_{,\phi \phi} \gamma^{-3} - a^2 \frac{h_{,\phi}}{2h} \frac{Q}{\rho_c} \left( 1 - 3 \gamma^{-2} \right). \nonumber &
\end{flalign}

\noindent We have verified that, in the limit where $h \phi'^2 \ll a^{2}$ and $\gamma \rightarrow 1$, we recover the disformal quintessence case, studied in Refs. \cite{vandeBruck:2015ida,Mifsud:2017fsy}. Also, from the expression of $\delta Q$, we see that the disformal scenario introduces a dependence on the scale $k$, through the first term of $ \mathcal{Q}_5$, in Eq. \eqref{q5}. This is a well-known feature of disformal couplings \cite{Zumalacarregui:2012us, vandeBruck:2015ida, Mifsud:2017fsy}, and we expect it to be reflected in the growth and distribution of perturbations. For the DBI scalar field, and considering adiabatic perturbations, we have

\begin{equation}
    c_{s,\phi}^2 \equiv \left( \frac{\partial p}{\partial X} \right) \left( \frac{\partial \rho}{\partial X} \right)^{-1} = \frac{1}{\gamma^2} \leq 1,
\end{equation}

\noindent which is always positive, granting the perturbations free from instabilities. The fact that this model allows for $c_{s,\phi}^2 \neq 1$ may give rise to distinctive signatures, namely, at the level of the cosmic microwave background temperature anisotropies and the matter power spectrum, two of the perturbed observables we will be interested in studying.

The set of dynamical perturbed Einstein equations, Eqs. \eqref{peq1}-\eqref{peq4}, the perturbed continuity and Euler equations, Eqs. \eqref{conseu}-\eqref{CDM2}, and the perturbed equation of motion for the scalar field, Eq. \eqref{PerturbKG}, can be evolved numerically, under particular assumptions and initial conditions, to give different cosmological regimes of this theory.

\subsection{The growth of perturbations and the effective gravitational constant between dark matter particles}

In what follows, we derive the equation of motion for the density contrast in the subhorizon limit ($ k \gg a H$), a first analytical approximation from which we can extract some relevant features. In doing so, we apply the quasistatic approximation, where the time dependence of the gravitational potential is given through the matter and field perturbations, so that we may neglect time derivatives of the perturbations and metric potential, such as $\Phi'$ in Einstein's equations and in the expression for the perturbation of the coupling. Likewise, we neglect $\delta\phi''$ and $\delta\phi'$ in the Klein--Gordon equation, assuming these are negligible when compared to the other terms. Thus, Eqs. \eqref{CDM1} and \eqref{CDM2} read, respectively,

\begin{eqnarray} \label{one}
\delta_c' &\simeq& -\theta_c - \frac{Q}{\rho_c}\phi'\delta_c + \frac{\delta Q}{\rho_c}\phi'  ,
\end{eqnarray}

\begin{eqnarray} \label{two}
\theta_c' &\simeq& - \mathcal{H} \theta_c + k^2\Psi - \frac{Q\phi'}{\rho_c}+k^2\frac{Q}{\rho_c}\delta\phi .
\end{eqnarray}

From the perturbed Einstein equations, we extract an approximation for the Poisson equation for the gravitational field, ignoring the contribution of baryons and radiation, residual at the present. Thus, we have (note that $\Phi=\Psi)$

\begin{equation}\label{three}
    k^2\Psi \simeq - 4\pi G_N \rho_c \delta_c.
\end{equation}

\noindent Finally, the Klein--Gordon equation reads

\begin{equation}\label{four}
    \mathcal{A} \Psi + \left( k^2\gamma^{-2} + a^2 m_{\rm eff}^2 \right) \delta \phi + a^2\gamma^{-3}\delta Q \simeq 0, 
\end{equation}

\noindent where

\begin{eqnarray}
\mathcal{A} = \left[ 6 \mathcal{H} \left( 1- \gamma^{-2} \right) \phi'  - \frac{h_{,\phi}}{h^2} a^2  \left( 2 - 3 \gamma^{-1} + \gamma^{-3} \right) + a^2  \left( V_{,\phi} + Q \right) \left(3\gamma^{-1} -\gamma^{-3} \right) \right]
\end{eqnarray}

\noindent and 

\begin{eqnarray}
a^2 m_{\rm eff}^2 &=& -\frac{3h_{,\phi}}{h} \mathcal{H} \left( 1 - \gamma^{-2} \right) \phi' + \frac{h_{,\phi \phi}}{2h^2} a^2 \left( 1 - 3\gamma^{-2} + 2\gamma^{-3} \right) - \frac{3}{2} \frac{h_{,\phi}}{h} a^2 \left( V_{,\phi} + Q \right) \left( \gamma^{-1} - \gamma^{-3} \right)  \nonumber \\
&+& a^2 V_{,\phi \phi} \gamma^{-3} + \frac{h_{,\phi}^2}{2 h^3} a^2 \left( 1 - 3\gamma^{-1} + 3\gamma^{-2} - \gamma^{-3} \right).
\end{eqnarray}

Applying these approximations to the expression for the perturbation of the coupling, Eq. \eqref{deltaQ}, we have

\begin{equation}
    \delta Q \simeq \frac{a^{-2} \rho_c}{\gamma^{-2} + h \rho_c \gamma^{-3}} \left( \mathcal{Q}_1\delta_c + \mathcal{Q}_5 \delta \phi \right).
    \label{deltaqaprox}
\end{equation}

In the expression for the $\mathcal{Q}_5$ coefficient, Eq. \eqref{q5}, the $k^2$ term is dominant. Following the same procedure for Eq. \eqref{deltaqaprox}, we arrive at the very simple expression, in the subhorizon limit:

\begin{equation}
    \delta Q \simeq Q \delta_c,
\end{equation}

\noindent which we have verified numerically and also holds in other theories with conformal and disformal couplings \cite{Zumalacarregui:2012us, vandeBruck:2015ida, Mifsud:2017fsy}.
% In Fig. \ref{fig:deltaQandGeff} (upper plots) we plot the evolution of $\delta Q$ together with $Q\delta_c$ for the Models 1 and 2 in Table \ref{tabel1}. For scales of the order $k = 0.1$Mpc$^{-1}$, we find that the approximation holds from $z \approx 10$. 

\begin{figure}
      \subfloat{\includegraphics[height=0.295\linewidth]{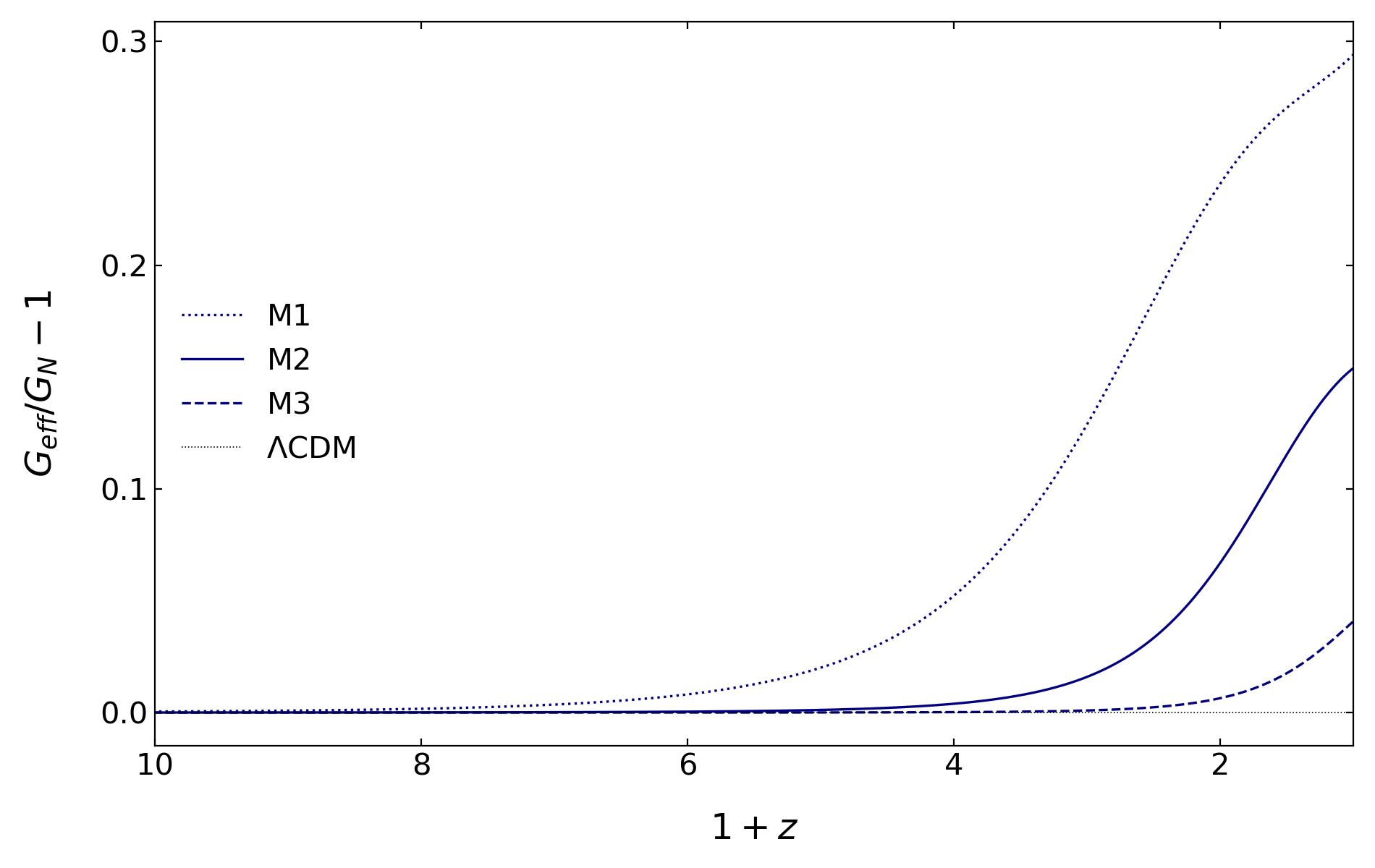}}
      \hfill
      \subfloat{\includegraphics[height=0.295\linewidth]{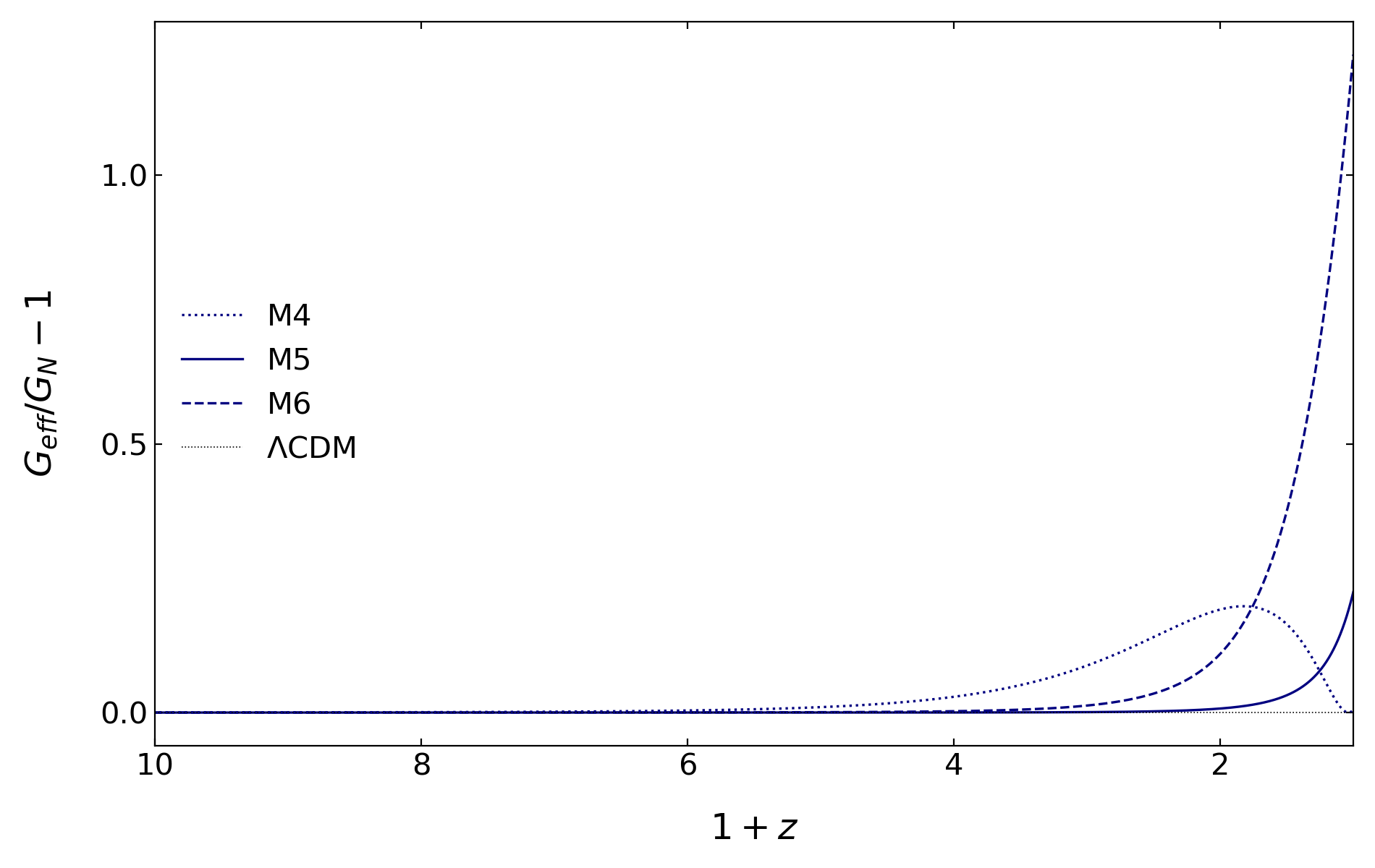}}
  \caption{\label{fig:Geff} Evolution of the effective gravitational constant, defined in Eq. \eqref{geff}, as function of the redshift $z$. The left and right panels correspond to models M1-M3 and M4-M6 in Table \ref{tabel1}, respectively, and the line styles used are the same as in the background evolution, depicted in Figs. \ref{fig:difgamma0beta} and \ref{fig:difgamma0}.}
\end{figure}

Taking the time derivative of the first equation in Eq. \eqref{two}, and using Eqs. \eqref{two}-\eqref{four}, we derive the following expression for the evolution of the density contrast $\delta_c$:

\begin{eqnarray}
\delta_c'' + \mathcal{H}_{\rm eff} \delta_c' \simeq  4\pi G_{\rm eff} \rho_c \delta_c,
\end{eqnarray}

\noindent where we have defined the effective Hubble rate

\begin{eqnarray}
   \mathcal{H}_{\rm eff} =  \left(\mathcal{H} + \frac{Q}{\rho_c}\phi'\right)
   \label{heff}
\end{eqnarray}

\noindent and the effective gravitational constant, given by 

\begin{eqnarray}
G_{\rm eff} = G_N\left(1 + 2\beta^2 \frac{1}{\gamma \left(1 + a^2m_{\rm eff}^2/k^2\gamma^2\right)} - A\frac{Q}{\rho_c}\frac{\gamma^2}{k^2} \right).
\end{eqnarray}

\noindent In the limits $k^2\gg a^2 m_{\rm eff}^2/\gamma^2$ and $k^2\gg A Q\gamma^2/\rho_c$, $G_{\rm eff}$ reduces to

\begin{eqnarray}
G_{\rm eff} \simeq G_N\left(1 + \frac{2 \beta^2}{\gamma} \right)
\label{geff}
\end{eqnarray}

\noindent with $\beta$ as defined in Eq. \eqref{beta}. $G_{\rm eff}$ is the effective gravitational coupling between two dark matter particles, comprised by standard gravity plus the long-range force mediated by the scalar field (second term in Eq. \eqref{geff}). 
Our result is consistent with that of \cite{Tsujikawa:2007gd} for scalar-tensor gravity models under a conformal transformation, generalised here to the disformal case, with a different functional form for $\beta$.

The evolution of $G_{\rm eff}/G_N$ for the models in Table \ref{tabel1} is shown in Fig. \ref{fig:Geff}. We see that $G_{\rm eff}$ deviates from $G_N$ at low redshift, whereas it tends to $G_N$ at early times. Consistently with what was found for $\beta$ in the background study, we see that smaller values of $\phi_{\rm ini}$ lead to higher values of $G_{\rm eff}$ at the present. On the other hand, in this case, the coupling also turns on later, and, therefore, interestingly, we expect that there will be fewer variations at the level of the CMB anisotropies and the matter power spectrum, when compared to $\Lambda$CDM, similar to what was observed in \cite{vandeBruck:2017idm} for disformally coupled models with a standard kinetic term. The early growth of $G_{\rm eff}$ is potentially problematic for the viability of the theory, when compared to observations. However, we also show an example in which the effective coupling starts out by being significantly positive but eventually starts decreasing (model M4 in the right panel in Fig. \ref{fig:Geff}), becoming smaller at the present. This means that there are parameter choices for which $G_{\rm eff} \simeq G_N$ today, but structure formation is influenced by the scalar-field-mediated force at intermediate redshifts. It is clear that the different evolution of $G_{\rm eff}$ in each model will translate into different behaviours for the matter fluctuations. This is the focus of the next subsection.

\subsection{CMB anisotropy and matter power spectra}

\begin{figure}
       \subfloat{\includegraphics[height=0.29\linewidth]{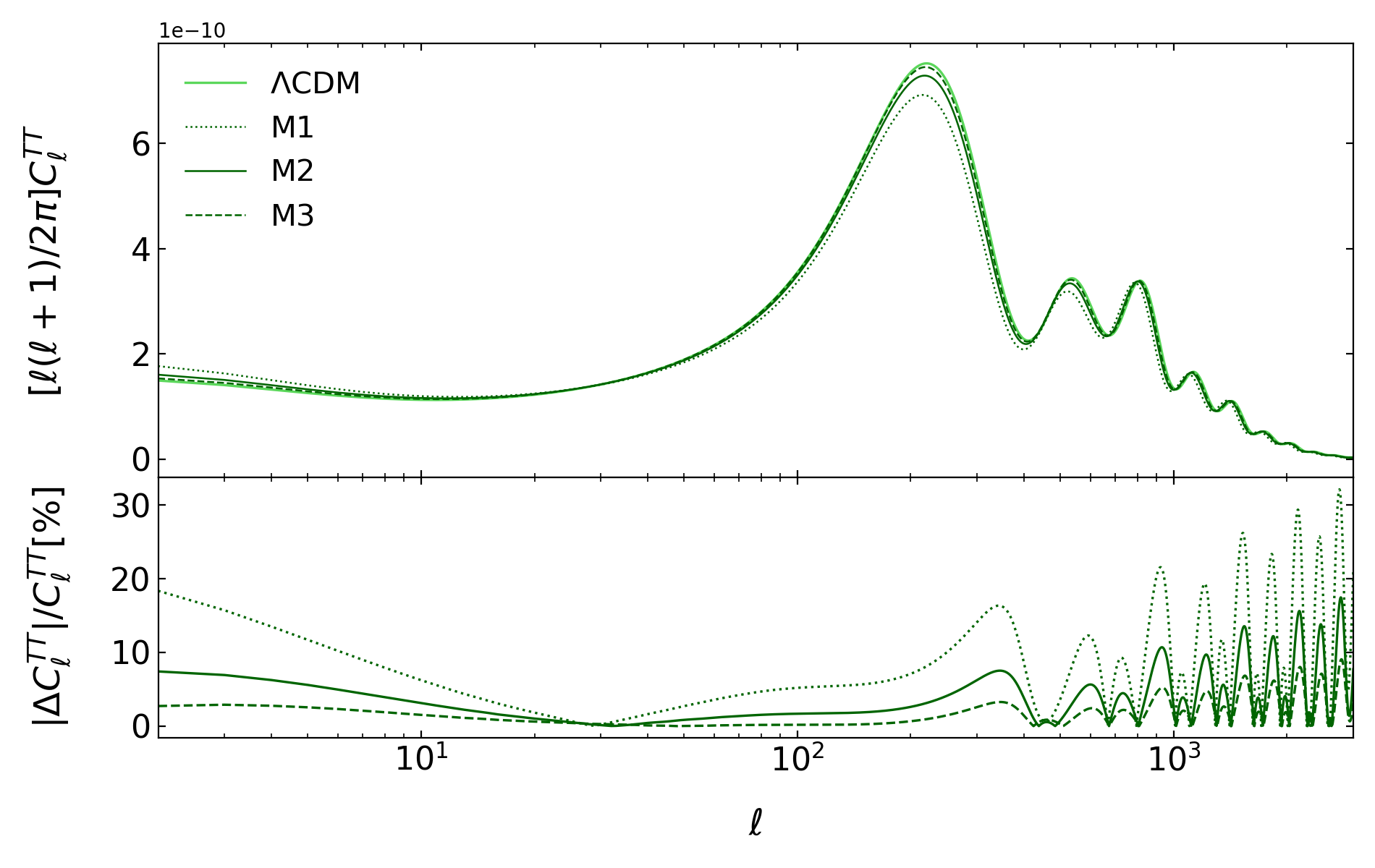}} 
       \hfill
       \subfloat{\includegraphics[height=0.29\linewidth]{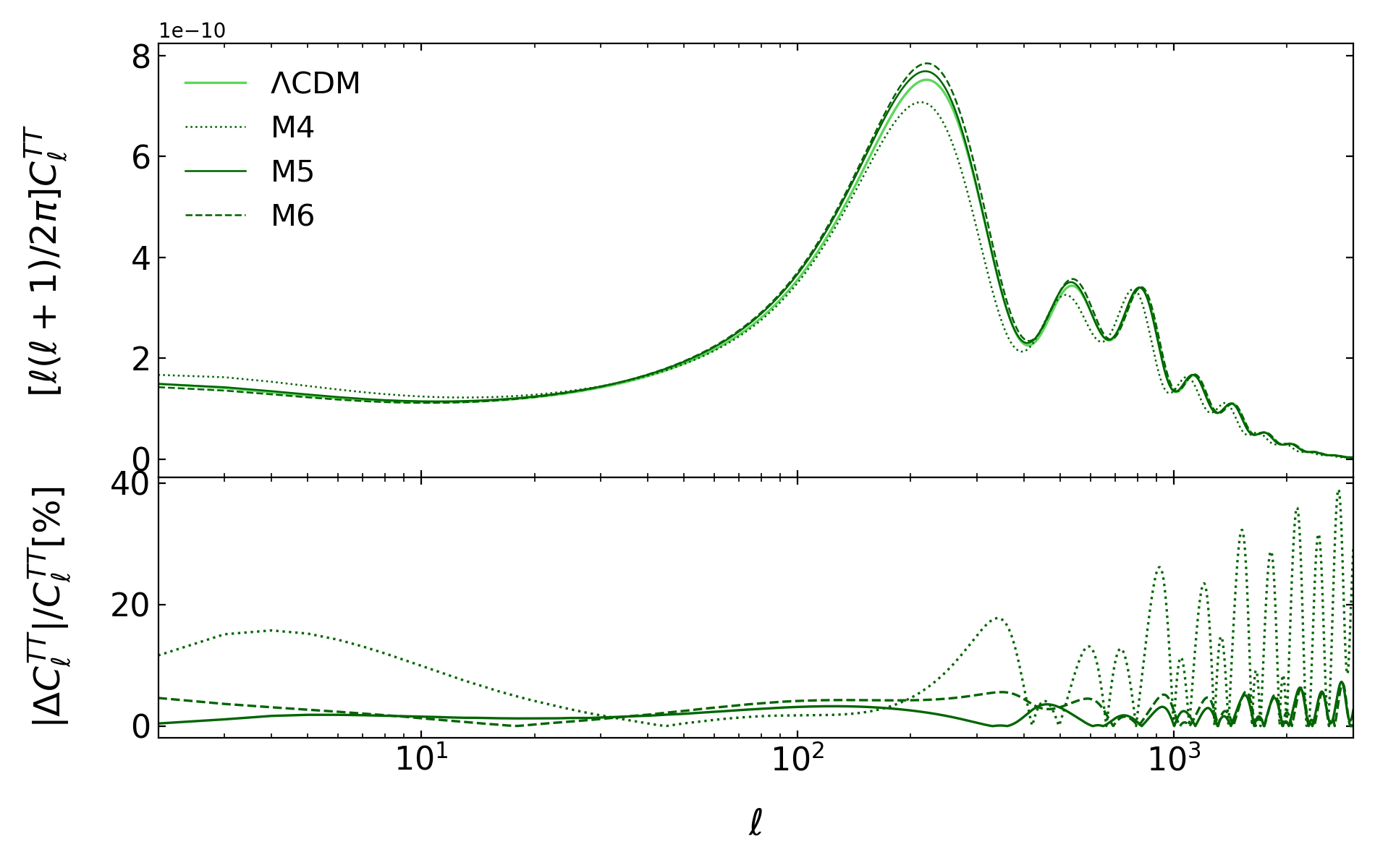}} 

      \subfloat{\includegraphics[height=0.29\linewidth]{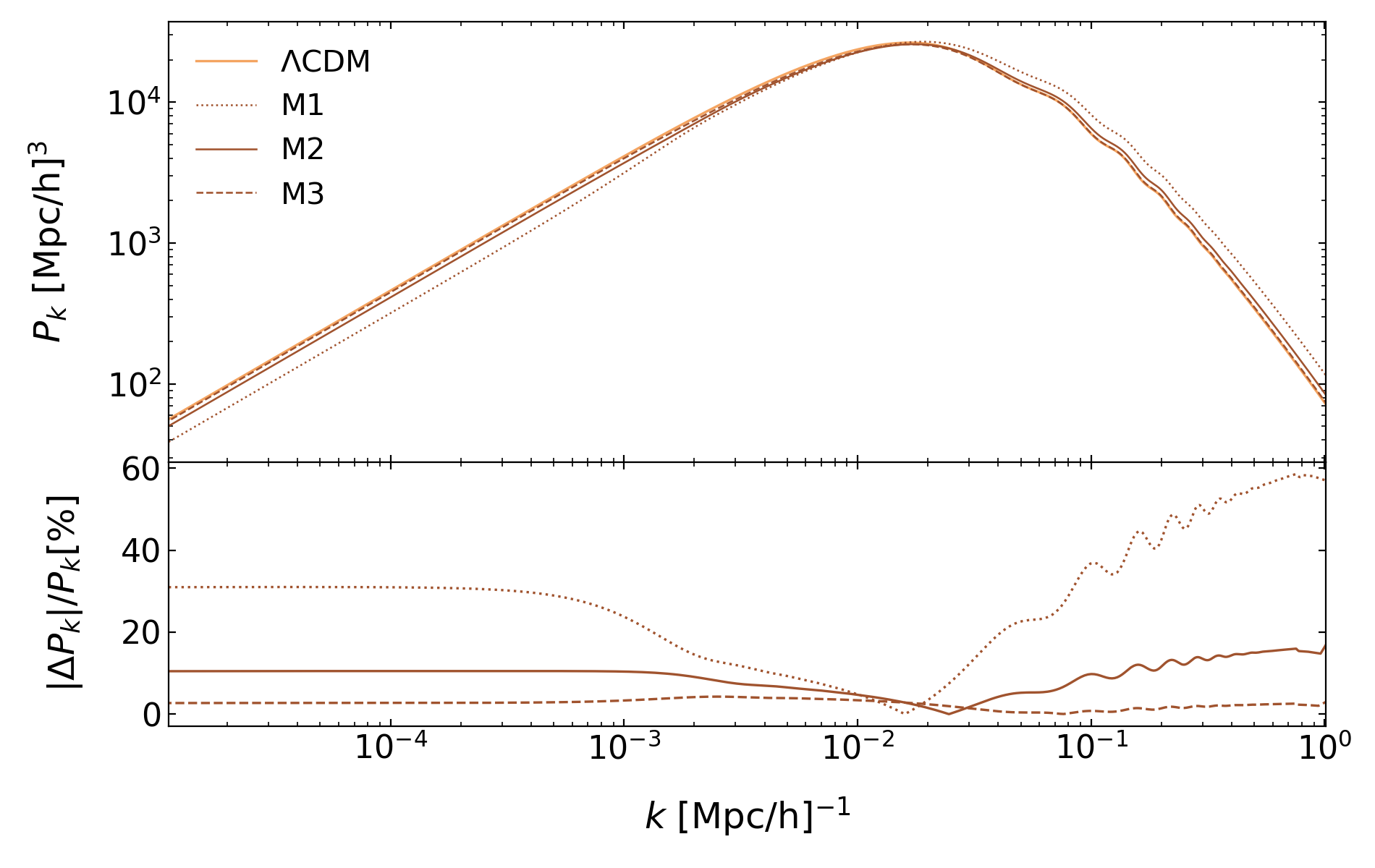}}
             \hfill
        \subfloat{\includegraphics[height=0.29\linewidth]{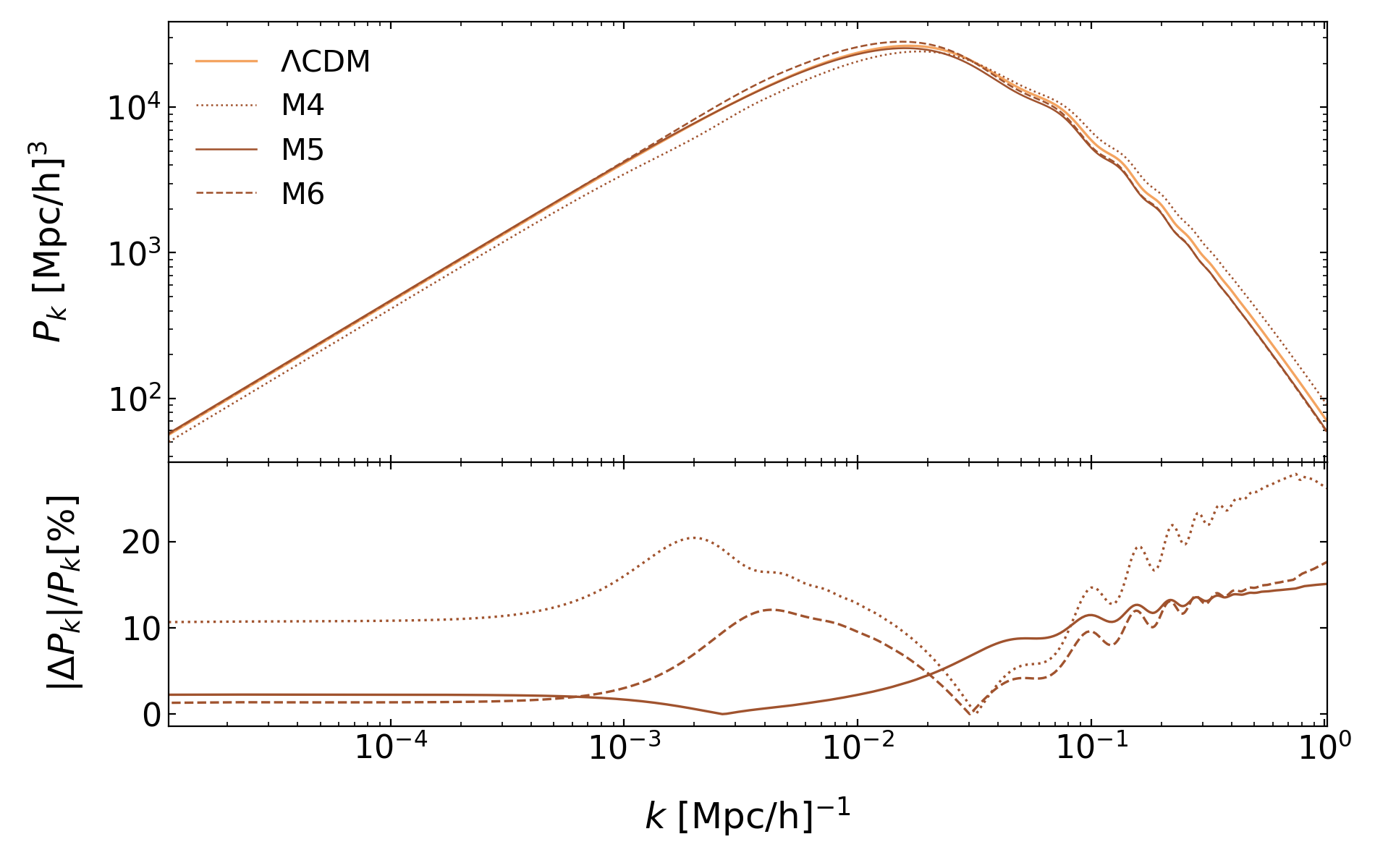}}
             \hfill     
  \caption{\label{fig:Powerspectra} The top panels show the CMB temperature-temperature angular power spectrum, $C_{\ell}^{TT}$, plotted as a function of the angle scale $\ell$. The bottom panels depict the matter power spectra $P_k$, for different Fourier scales (wave numbers) $k$. For each observable, we also provide the $\Lambda$CDM predictions (lighter shades), together with the relative deviation of each model from the standard model. The left and right panels correspond to models M1-M3 and M4-M6 in Table \ref{tabel1}, respectively, and the line types used are the same as in Figs. \ref{fig:difgamma0beta} and \ref{fig:difgamma0} for the background evolution.}
\end{figure}

To evaluate the CMB anisotropy power spectrum and the matter power spectrum in the dark D-brane scenario, we resorted to the CLASS code \cite{Lesgourgues:2011re, Blas:2011rf}, modified for our purposes. The results for the cases in Table \ref{tabel1}, together with the predictions for the $\Lambda$CDM model, are depicted in Fig. \ref{fig:Powerspectra}, in which we present the CMB temperature anisotropies (top panels) and the matter power spectra for DDM and baryons at $z = 0$ (bottom panels). In all cases, we have considered standard adiabatic initial conditions with an amplitude $A_s  = 2.215 \cdot 10^{-9}$ and $k_{\rm pivot}=0.05 {\rm Mpc}^{-1}$, and vanishing initial perturbations for the scalar field, $\delta \phi_{\rm ini} = \delta \phi'_{\rm ini} = 0$. Similarly to the background cosmology, the linear perturbations feature a strong dependence on the parameters. This is confirmed in Fig. \ref{fig:Powerspectra}, where different choices for $\Gamma_0$, but fixed $\phi_{\rm ini}$, result in different predictions for the power spectra. On the other hand, in Fig. \ref{fig:Powerspectra} we also present two distinct solutions for the same values of $\Gamma_0$, one for $\phi_{\rm ini} = 3$ M$_{\rm Pl}$ (models M1-M3 on the left panel) and another $\phi_{\rm ini}  =1.7$ M$_{\rm Pl}$ (models M4-M6 on the right panel), characterised by the same line styles. As expected, different initial field values also result in distinct cosmological imprints at the level of linear perturbations.

By comparing the left and right panels in Fig. \ref{fig:Powerspectra}, one concludes that the sign of the effective coupling, $\beta$, which depends on $\Gamma_0$ and $\phi_{\rm ini}$, has a clear effect on the evolution of the perturbations. For the CMB anisotropy power spectrum, we observe that, for $\beta>0$ ($<0$), the curves are generally above (below) the $\Lambda$CDM case for low $\ell$ and below (above) for medium and high values of $\ell$. Also, when $\beta>0$ ($<0$), higher (lower) values of $\Gamma_0$ result in fewer deviations from the $\Lambda$CDM model. It is interesting to note that, even though the effective coupling today is much smaller in the models M1-M4 (see Fig. \ref{fig:difgamma0beta}), the deviations from $\Lambda$CDM are much larger when compared to M5 and M6. This can be ascribed to the fact that the coupling becomes significant at higher redshifts, leading to earlier deviations from the standard model. 

The reason why the predictions for the CMB anisotropies in the models differ is twofold: First, the background evolution is not the same, as explained in Sec. \ref{sec:back}. In particular, the ratio $\Omega_b/\Omega_{\rm c}$ is, in general, not constant, and its evolution and value at high redshifts depend on the parameters chosen. This is the main contributor to the shifts in position and changes in amplitude of the various peaks, which look narrower (wider) for models M1-M4 (M5 and M6), when compared to the $\Lambda$CDM case. Second, the evolution of $G_{\rm eff}$, directly related to the coupling, $\beta$, and the Lorentz factor, $\gamma$, influences the growth of perturbations in the DDM fluid. In the case of models M1-M4, the effective gravitational coupling $G_{\rm eff}$ may start to grow at a redshift as high as about $6$, for the values of $\Gamma_0$ considered, whereas for models M5 and M6, $G_{\rm eff}$ starts to grow only at a redshift of about 2 or later. The growth of the effective coupling at late times influences the late time integrated Sachs--Wolfe (ISW) effect. 

\begin{figure}
      \subfloat{\includegraphics[height=0.295\linewidth]{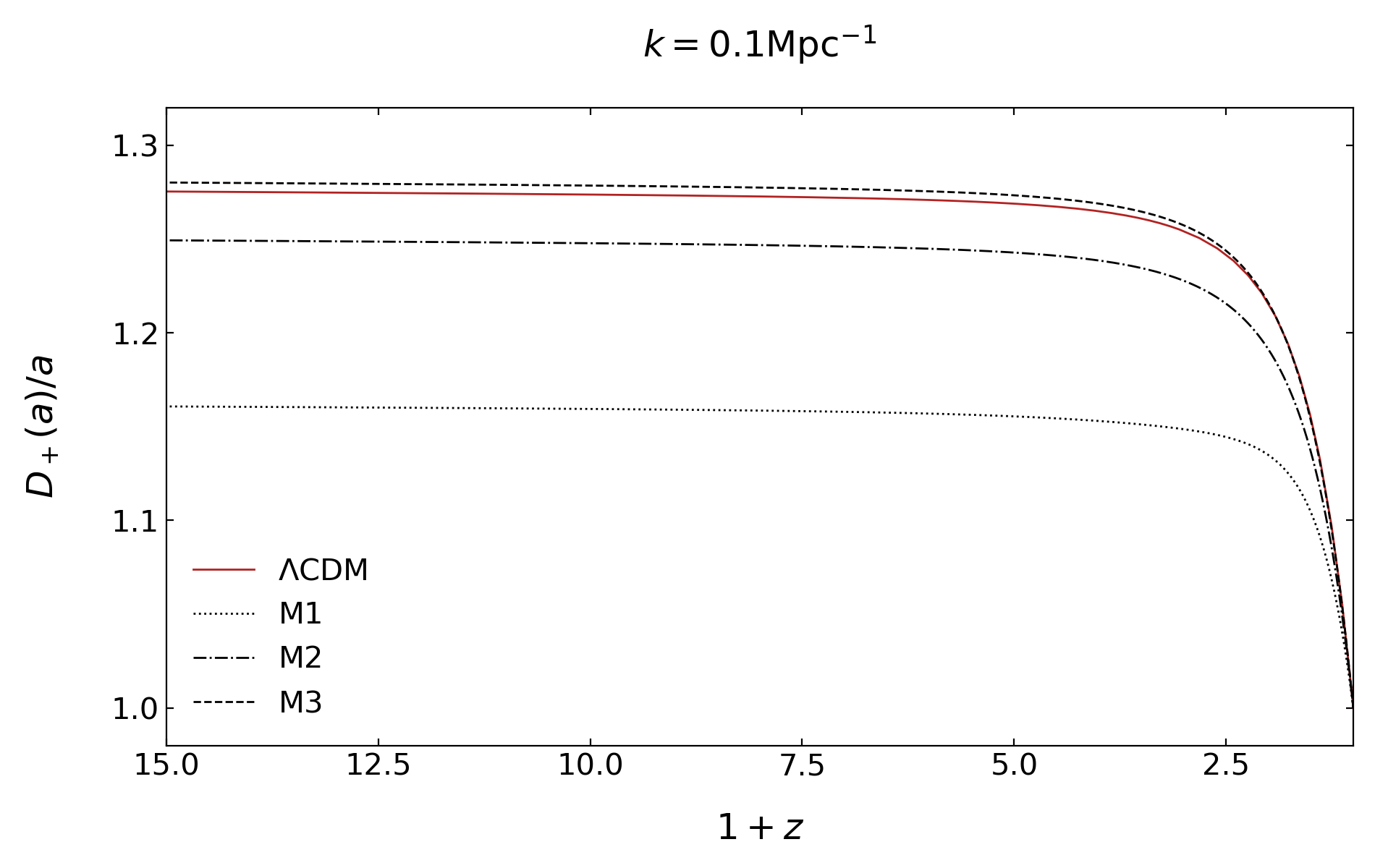}}
             \hfill
        \subfloat{\includegraphics[height=0.295\linewidth]{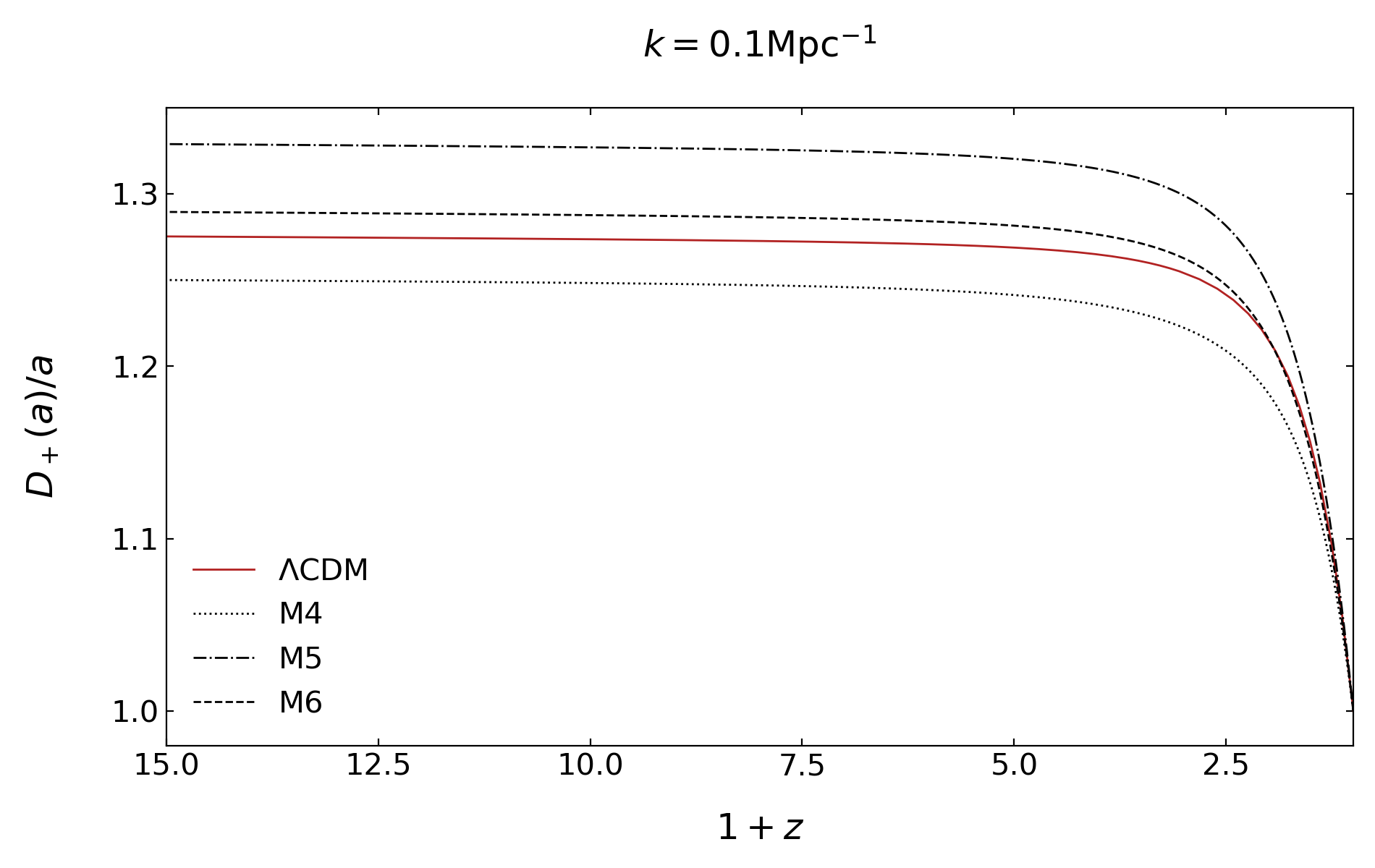}}
             \hfill     
  \caption{\label{fig:growthfactor} Evolution of the growth factor $D_+$, normalised according to its value at $z=0$ and divided by the scale factor $a$. The plots depict the predictions for $k=0.1$Mpc$^{-1}$, plotted as a function of the redshift $z$. The left and right panels correspond to models M1-M3 and M4-M6 in Table \ref{tabel1}, respectively, and the line styles used are the same as in Figs. \ref{fig:difgamma0beta} and \ref{fig:difgamma0} for the background evolution.}
\end{figure}

\begin{figure}
    %   \subfloat{\includegraphics[height=0.285\linewidth]{deltam43.png}} 
    %   \hfill
    %   \subfloat{\includegraphics[height=0.285\linewidth]{deltam41p7.png}} 

      \subfloat{\includegraphics[height=0.295\linewidth]{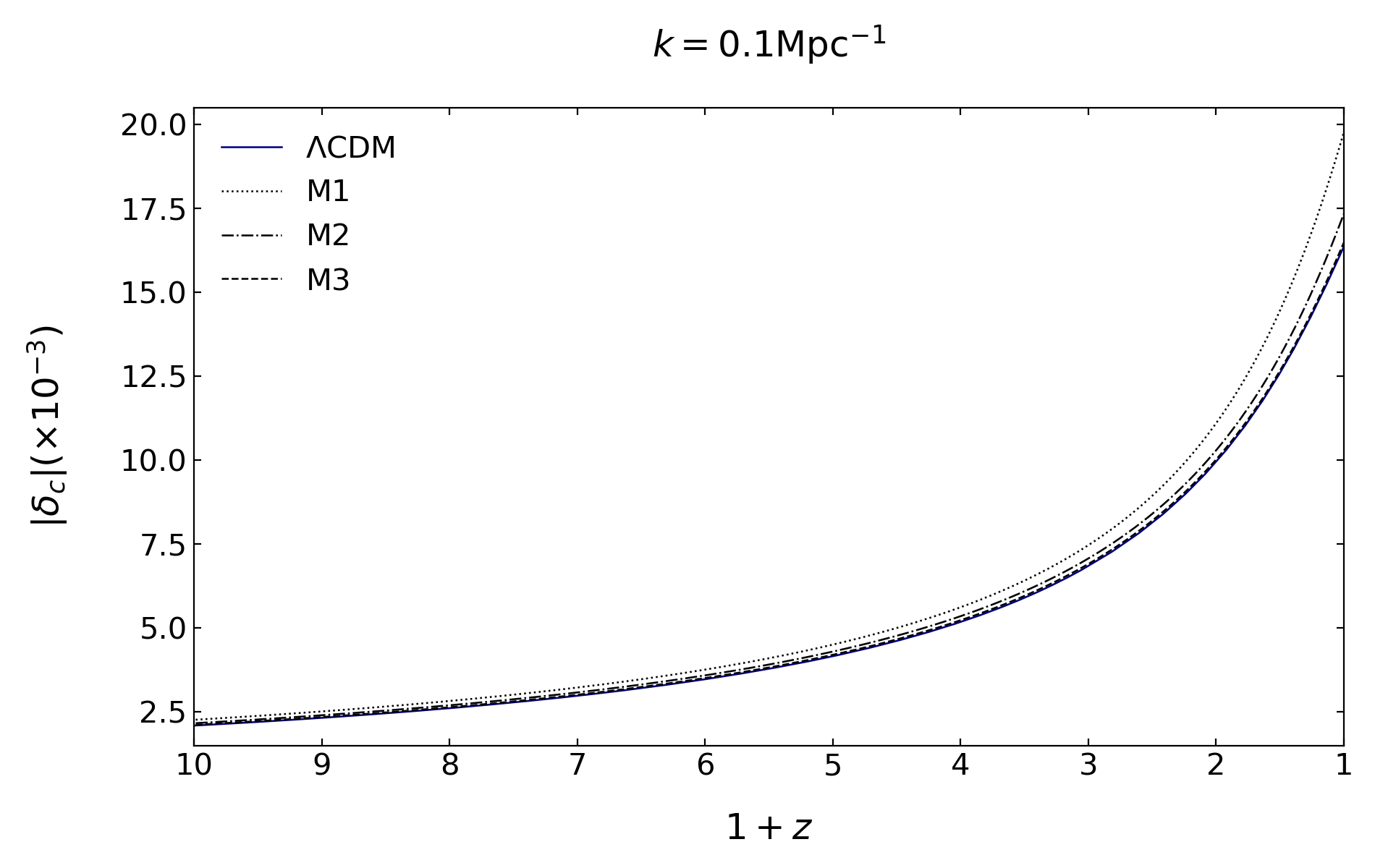}}
             \hfill
        \subfloat{\includegraphics[height=0.295\linewidth]{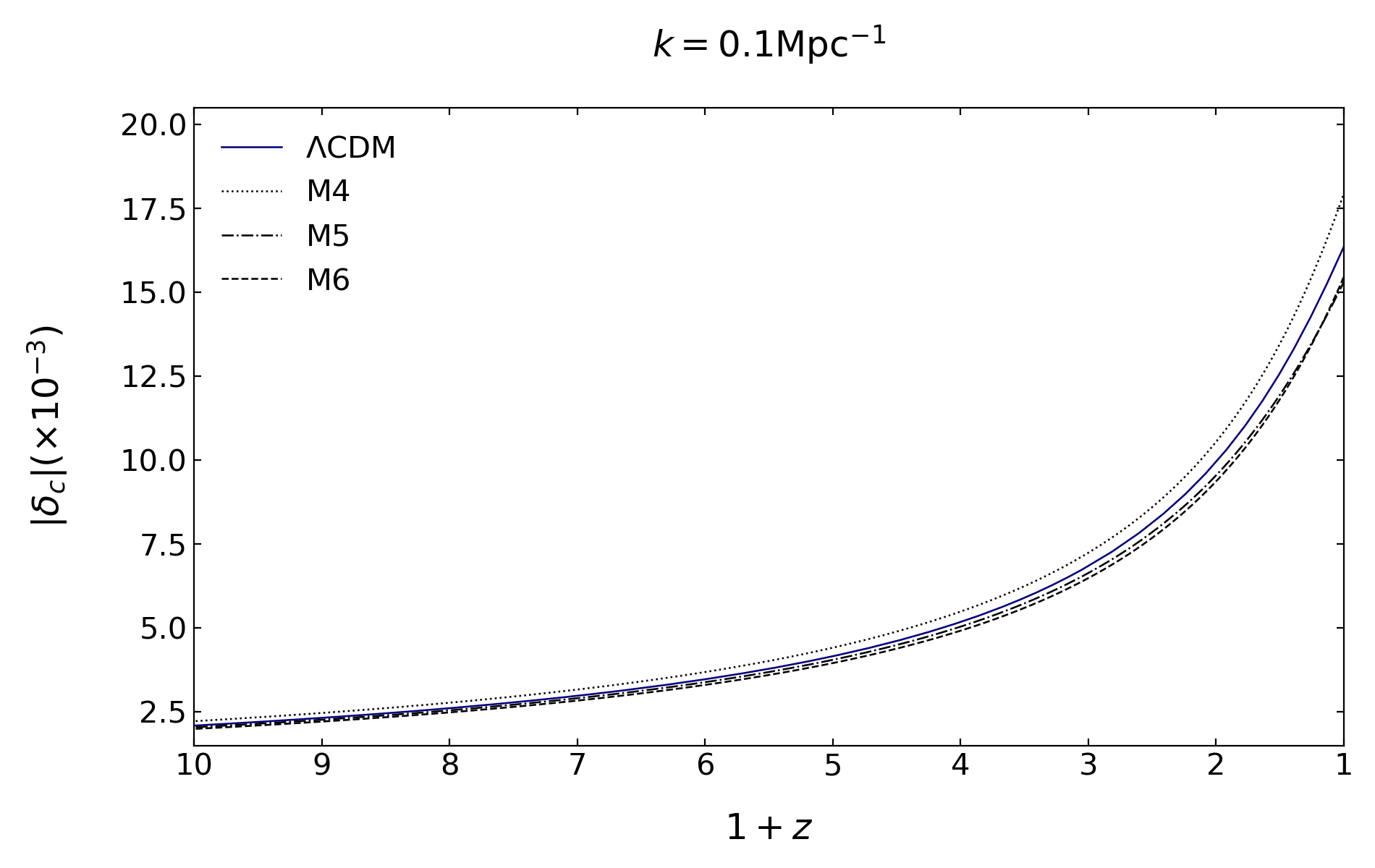}}
             \hfill     
  \caption{\label{fig:Deltas} Evolution of the density contrast of disformal dark matter, $\delta_c$, for the scale $k=0.1$Mpc$^{-1}$, plotted as a function of the redshift $z$. The left and right panels correspond to models M1-M3 and M4-M6 in Table \ref{tabel1}, respectively, and the line types used are the same as in Figs. \ref{fig:difgamma0beta} and \ref{fig:difgamma0} for the background evolution.}
\end{figure}

The predictions for the matter power spectrum, shown in Fig. \ref{fig:Powerspectra} for all cases, depend on various factors. Apart from the different background evolution, the change in the effective gravitational coupling $G_{\rm eff}$, defined in Eq. \eqref{geff}, and the effective Hubble expansion rate, defined in Eq. \eqref{heff}, have a major influence on the evolution of the dark matter density contrast, as discussed in the last subsection. 
In the bottom panels in Fig. \ref{fig:Powerspectra} we see that, for models M1-M4 (M5 and M6), on very large scales, $k \lesssim k_{\rm peak}$, the dynamical evolution of dark energy contributes to a suppression (enhancement) of the growth of structures, when compared to the $\Lambda$CDM case, whereas on small scales, $k \gtrsim k_{\rm peak}$, the opposite holds. In agreement with the discussion for the CMB power spectrum, the deviations from the standard model are more pronounced for models M1-M4, for which the coupling is activated at earlier redshifts.
In addition, at least for some period of time, the perturbation of the coupling, $\delta Q$, is scale dependent (through the coefficient ${\cal Q}_5$, Eq. \eqref{q5}).
The predicted baryon oscillations are also clearly imprinted on the matter power spectrum at small scales.
To quantify the effect of the modified growth of density perturbations, we plot in Fig. \ref{fig:growthfactor} the evolution of the quantity $D_+/a$, where $D_+ = \delta_{\rm M}/\delta_{M,0}$ ($M$ stands for total nonrelativistic matter: dark matter and baryons) is the growth factor, normalised at $z=0$, for the cases presented in Table \ref{tabel1} and the $\Lambda$CDM model. At high redshift, the difference to the concordance model can be as high as about $10\%$ for M1 and can be almost negligible ($\lesssim 1\%$) for models M3 and M6. With the normalisation chosen, for models M1-M4, the amplitude of primordial perturbations will have to be larger to obtain the same number of structures today, with the opposite holding for M5 and M6. This is in agreement with what is presented in Fig. \ref{fig:Deltas} for models M1-M4, where we show the evolution of the density contrast $\delta_{c}$ of the DDM fluid for the scale $k = 0.1 $Mpc$^{-1}$. This is also consistent with the fact that these models show a slightly higher dark energy fractional density parameter at late times, when compared to the $\Lambda$CDM case (see red curves in the left upper panel in Fig. \ref{fig:difgamma0}). The early onset of dark energy suppresses the growth of perturbations and acts against the growing influence of the coupling. In models M5 and M6 dark energy becomes important at a later redshift, when compared to M1-M4. At early times, when the presence of dark energy is negligible, the growth rate is proportional to the scale factor for all models: $D_+ (a) \propto a$. 

The ISW effect is proportional to $d(D_+(a)/a)/da$. For models M1-M4, the derivative of $D_+(a)/a$ is larger (in absolute value) at late times, and, therefore the ISW effect will be more pronounced, when compared to the $\Lambda$CDM case, giving rise to the enhancement of the low-$\ell$ tail of the CMB anisotropy spectrum depicted in the top panels of Fig. \ref{fig:Powerspectra}, with the opposite holding for M5 and M6. Another contribution to the ISW effect would be a period of early dark energy. However, in contrast to other dark energy models with disformal couplings, (\textit{e.g.}, \cite{Zumalacarregui:2012us}), we do not observe any significant early dark energy signatures in the dark D-brane scenario, at least not with the form of the potential we have chosen in this work.

From Table \ref{tabel1}, we also note that models M1-M4 (M5 and M6) predict values of $\sigma_8$ that are above (below) the one for the $\Lambda$CDM model, that is, $\sigma_8^{\Lambda {\rm CDM}}= 0.848$.

The DDM scenario shares some similarities with other disformal models discussed in the literature. As reported in previous works, the effective gravitational coupling between dark matter particles is not constant in disformal models \cite{Zumalacarregui:2012us,vandeBruck:2015ida,Mifsud:2017fsy,vandeBruck:2017idm} (whereas, \textit{e.g.}, in the standard conformally coupled quintessence scenario, it is often chosen to be constant; see \cite{Wetterich:1994bg,Amendola:1999er}). In the dark D-brane setting, the coupling is negligible in the very early Universe, because it is suppressed by the denominator in Eq. \eqref{beta}. This is yet another model which motivates searches for violations of the equivalence principle in the dark sector at late times. In future work, we will study in more depth the impact of the time-evolving coupling on structure formation and compare the DDM model to cosmological data \cite{Teixeira_future}.

\section{Conclusions} \label{sec:conc}

In this work, we have analysed the evolution of the background and linear cosmological perturbations for the dark D-brane model, introduced in Ref. \cite{Koivisto:2013fta}. To probe the effects of a disformal coupling between the dark energy component and dark matter residing on the D-brane, we have chosen six sets of parameters, which allow us to study two regimes, distinguished by the behaviour of the coupling at different redshifts. The coupling is always small initially and starts to grow at intermediate redshifts only. In one case, the coupling is positive throughout the evolution. In the other, the coupling may start out as being positive, before becoming smaller and approaching negative values at late times. These two classes of models introduce distinct and rich phenomenology for the evolution of both the background and the linear perturbations. 

At the background level, and near the fixed point solutions, we found that the model resembles $\Lambda$CDM for increasingly higher values of the parameter $\Gamma_0$, defined in Eq. (\ref{Gamma0}), as expected from the dynamical systems study. In all models, the Universe is evolving towards a state characterised by a scaling fixed point, where the fractional energy densities of DE and DM maintain a constant ratio. After spending a certain number of \textit{e}-folds around this solution, the system will inevitably evolve towards a state where the dynamics of the Universe is fully dominated by the DBI scalar field. 
This model is devoid of phantomlike behaviour, as the equation of state parameter for dark energy is always larger, albeit very close, than $-1$. However, the effective EoS parameter, defined in Eq. \eqref{weffphi}, can mimic an apparent phantom nature, by transposing the effects of the coupling between DM and DE to an effective noninteracting dark energy fluid. In addition to the dynamical system analysis conducted in Ref. \cite{Koivisto:2013fta}, we find that the initial condition of the field plays an important role, since this quantity also affects the evolution of the coupling. By fixing $\Gamma_0$, we gather that, by taking initial conditions for the field that lead to higher present values of $\beta$, the coupling starts to grow much earlier, during the matter-dominated epoch, leaving distinct signatures at the level of the perturbations. Therefore, we conclude that the difference in the direction of the energy exchange between the dark fluids introduces distinct features at the level of the background, potentially shifting the matter-radiation equality and the time of transition from matter- to dark energy- dominated epochs.

In addition to studying the evolution of the background, we have derived the equations for the evolution of cosmological perturbations in both the Newtonian and synchronous gauges and calculated the resulting CMB anisotropies power spectrum and matter power spectrum. One of the distinct features of the DBI models is the fact that the sound speed depends on the Lorentz factor, and, therefore, it will generally deviate from the speed of light. As expected, the two different background regimes also lead to  distinct signatures in the growth of the perturbations. For the CMB power spectrum, we identify an enhancement or suppression of the ISW tail at low multipoles, with the deviations becoming higher for the models with the greater effective coupling today, along with a slight shift and change in shape of the acoustic peaks at high multipoles. We find that the deviations from $\Lambda$CDM are actually larger for models M1-M4, with smaller couplings at the present. The fact that the coupling turns on earlier has a great impact on the time evolution of the perturbations and is also reflected on the larger deviations of the matter power spectrum of models M1-M4, when compared to the $\Lambda$CDM model. 

In our simulations, the initial value of the scalar field was always chosen to be of the order of the Planck mass. In terms of the higher-dimensional picture, this means that the D--brane is far away from the tip of the AdS throat. In Ref. \cite{Koivisto:2013fta}, another limit was discussed, namely, the region in which $\phi \ll 1$ M$_{\rm Pl}$ and the brane is moving near the tip of the AdS throat. In this scenario, and in order to explain the observed value of the vacuum energy today, the mass of the scalar field, $m_{\phi}$, is constrained only by a lower bound, and, therefore, does not have to be necessarily small, in contrast to standard quintessence models. However, we found that this scenario is challenging to study numerically, since, in addition to a very small initial condition for the scalar field, large values of $\Gamma_0$ need to be considered. It would be of interest to study this situation in more detail in the future. Nevertheless, we expect that the physics presented in this work should remain relevant in this regime.

In general, the numerical results presented in this paper hint at the possibility of finding distinct observational imprints of the dark D-brane model in the CMB and large scale structures. In future work, we will perform a Markov Chain Monte Carlo (MCMC) analysis to constrain the parameters of the theory \cite{Teixeira_future} according to current observational data. Some models of coupled dark energy alleviate existing tensions among the cosmological parameters obtained from various datasets \cite{DiValentino:2019jae,Gomez-Valent:2020mqn}, and, therefore, it remains to be seen whether the dark D--brane model belongs to such a class of coupled models. Over the next decade, new cosmological data will be available from next-generation surveys, such as Euclid \cite{Laureijs:2011gra, Amendola:2012ys} and DESI \cite{Levi:2019ggs}. Since these observations are expected to introduce further constraints on theories beyond the $\Lambda$CDM scenario, a complete study of the observational signatures of cosmological models is of paramount importance. Nonstandard cosmological scenarios, such as the dark D-brane model or the ones studied in, \textit{e.g.}, Refs. \cite{Zumalacarregui:2012us,Mifsud:2017fsy,vandeBruck:2019vzd}, also motivate the searches for late-time equivalence principle violations signatures in the dark sector, for which a detailed study of the evolution of nonlinear perturbations is needed.

\acknowledgments 
CvdB is supported by the Lancaster-Manchester-Sheffield Consortium for Fundamental Physics under STFC grant ST/P000800/1. EMT is supported by the grant SFRH/BD/143231/2019 from Funda\c{c}\~ao para a Ci\^encia e a Tecnologia (FCT). 

\begin{appendices}

\section{Synchronous Gauge} \label{app:synch}

In the main text, we have worked in the Newtonian gauge, for simplicity of the analysis. When studying the evolution of perturbations and calculating the resulting power spectra, we also worked in the synchronous gauge. Therefore, for completeness, in this Appendix, we provide the perturbation equations in the synchronous gauge for a generic conformal coupling function, $C(\phi)$, and disformal coupling function, $D(\phi)$. for a coupled DBI field. We also particularise the equations to the case considered in this work, in which $C(\phi)$ and $D(\phi)$ are related to the warp factor $h(\phi)$. 

The line element in the synchronous gauge can be written as 

\begin{equation}
ds^2= a^2 \left( \tau \right) \left[-d\tau^2 + \left(\delta_{ij} + h_{ij} \right) dx^{i} dx^{j} \right],
\end{equation}

\noindent where $h_{i j}$ represents the metric perturbation. In what follows, we will adopt Ma-Bertschinger \cite{Ma:1995ey} notation, although we use $\mathpzc{h}$ and $\eta$ for the scalar metric perturbations instead of the original $h$ and $\eta$ in order to avoid any confusion with the warp factor $h(\phi)$. In what follows, we write down the relevant equations in Fourier space, with $k$ standing for the Fourier modes (wave numbers).

Analogously to the treatment for the Newtonian gauge, we derive the perturbed Einstein equations:

\begin{equation}
k^2 \eta - \frac{1}{2} \mathcal{H} \mathpzc{h}' = -4 \pi G_N a^2 \sum \delta \rho_f
\end{equation}

\begin{equation}
k^2  \eta' = 4 \pi G_N a^2 \sum \rho_f \left( 1 + w_f \right) \theta_f
\end{equation}

\begin{equation}
\mathpzc{h}'' + 2\mathcal{H} \mathpzc{h}' - 2 k^2 \eta = - 24 \pi G_N a^2 \sum \delta p_f
\end{equation}

\begin{equation}
\mathpzc{h}'' + 6 \eta'' + 2 \mathcal{H} \left( \mathpzc{h}' + 6 \eta' \right) - 2 k^2 \eta = - 24 \pi G_N a^2 \sum \rho_f \left( 1 + w_f \right) \sigma_f
\end{equation}

\noindent with $\theta_f= \nabla_i v_f^i$ and $\sigma_f = \frac{2 w_f \Pi_f}{3 \left( 1 + w_f \right)}$, with $v_f^i$ being the velocity perturbation and $\Pi_f$ the anisotropic stress.

The perturbed continuity and Euler equations for the uncoupled baryonic and radiation fluids are derived directly from the perturbation of the conservation relations $ \nabla_{\mu} T^{\mu}_{\nu}$. They are given by 

\begin{equation}
\delta_u' + 3 \mathcal{H} \left( c_{s,u}^2-w_u \right) \delta_u = - \left(1 + w_u \right) \left( \frac{\mathpzc{h}'}{2} + \theta_u \right)
\end{equation}

\begin{equation}
\theta_u' + \left[ \mathcal{H} \left( 1 - 3 w_u \right) + \frac{w_u'}{1+w_u} \right] \theta_u = k^2 \frac{c_{s,u}^2}{1 + w_u} \delta_u - k^2 \sigma_u,
\end{equation}

\noindent with $u=\{b,r\}$. For DDM, we need the coupled sheer-free version of these equations, presented below for a general equation of state and adiabatic sound speed: 

\begin{equation}
\delta_c' + 3 \mathcal{H} \left( c_{s,c}^2-w_c \right) \delta_c = - \left(1 + w_c \right) \left( \frac{\mathpzc{h}'}{2} + \theta_c \right) - \frac{Q}{\rho_c} \delta_c \phi' + \frac{Q}{\rho_c} \delta \phi' + \frac{\delta Q}{\rho_c} \phi'
\end{equation}

\begin{equation}
\theta_c' + \left[ \mathcal{H} \left( 1 - 3 w_c \right) + \frac{w_c'}{1+w_c} \right] \theta_c = k^2 \frac{c_{s,c}^2}{1 + w_c} \delta_c - \frac{Q}{\rho_c} \phi' \theta_c + k^2 \frac{Q}{\rho_c \left(1+w_c \right)} \delta \phi
\end{equation}

For the energy density and pressure of the scalar field, we also have 

\begin{equation}
\delta \rho_\phi = a^{-2} \gamma^{3} \phi' \delta \phi '  + \frac{h_{,\phi}}{2 h^2} \left( 2 - 3 \gamma + \gamma^3 \right) \delta \phi + V_{, \phi} \delta \phi,
\end{equation}

\begin{equation}
\delta p_\phi = a^{-2} \gamma \phi' \delta \phi ' - \frac{h_{,\phi}}{2 h^2} \left( 2 - \gamma^{-1} - \gamma \right) \delta \phi - V_{, \phi} \delta \phi.
\end{equation}

The perturbed Klein Gordon equation in the synchronous gauge reads 

% CHOOSE!!

% \begin{align}
% &\delta \phi'' + \frac{\mathpzc{h}'}{2} \gamma^{-2} \phi' + k^2 \gamma^{-2} \delta \phi - \mathcal{H} \left( 7 - 9 \gamma^{-2} \right) \delta \phi' - 3 h \gamma^{-1} Q \phi' \delta \phi' + a^2 \frac{3}{2}\frac{h_{, \phi}}{h} Q \left( \gamma^{-3} - \gamma^{-1} \right) \delta \phi  \nonumber \\
% & - 3 \frac{h_{, \phi}}{h} \mathcal{H} \left(1-\gamma^{-2} \right) \phi' \delta \phi + 3 \frac{h_{, \phi}}{h} \left(1-\gamma^{-1} \right) \phi' \delta \phi' + \frac{a^2}{2} \frac{h_{, \phi}^2}{h^3} \left(1- \gamma^{-3} + 3 \gamma^{-2} - 3 \gamma^{-1} \right) \delta \phi + a^{2} \gamma^{-3} \delta Q  \\
% &- 3 h \gamma^{-1} V_{, \phi} \phi' \delta \phi' - a^2 \frac{3}{2} \frac{h_{, \phi}}{h} V_{, \phi} \left( \gamma^{-1} - \gamma^{-3} \right) \delta \phi + \frac{a^2}{2} \frac{h_{, \phi \phi}}{h^2} \left( 1 + 2 \gamma^{-3} - 3 \gamma^{-2} \right) \delta \phi + a^{2} V_{, \phi \phi } \gamma^{-3} \delta \phi = 0 .\nonumber
% \end{align}

\begin{align}
&\delta \phi'' + \frac{\mathpzc{h}'}{2} \gamma^{-2} \phi' + \left[  2 \mathcal{H}  + \frac{3}{4} \frac{h_{, \phi}}{h} \phi' +3 \frac{Q}{\rho_c} \phi' \right] \delta \phi' + \left[ k^2 \gamma^{-2}  + \frac{3}{2} \frac{h_{, \phi}}{h} \mathcal{H} \left(1-\gamma^{-2} \right) \phi' \right. \nonumber \\
& \left.  - \frac{a^2}{2} \frac{h_{, \phi}^2}{h^3} \left(\frac{5}{4} + 4 \gamma^{-3} - \frac{21}{4} \gamma^{-2} \right) + \frac{a^2}{2}  \frac{h_{, \phi \phi}}{h^2} \left( 1 + 2 \gamma^{-3} - 3 \gamma^{-2} \right)+ a^{2} V_{, \phi \phi } \gamma^{-3} \right] \delta \phi + a^{2} \gamma^{-3} \delta Q = 0,
\end{align}

\noindent with the perturbation of the coupling function $Q$ defined as 

\begin{equation}
\delta Q = \frac{a^{-2} \rho_c}{C - \frac{D}{h} \left( 1- \gamma^{-2} \right) + D \gamma^{-3} \rho_c} \left( \mathcal{Q}_1 \delta_c + \mathcal{Q}_2 \mathpzc{h}' + \mathcal{Q}_3 \delta \phi' + \mathcal{Q}_4 \delta \phi \right),
\end{equation}

\noindent with

% CHOOSE!!

% \begin{flalign}
% \mathcal{A}_1 = & \frac{1}{2} a^2 C_{,\phi} \left( 1 - 3 \frac{\delta p_c}{\delta \rho_c} \right) - 3 D \mathcal{H} \left( \gamma^{-2} + \frac{\delta p_c}{\delta \rho_c} \right) \phi' - a^2 D \gamma^{-3} \left( V_{, \phi} + Q \right) - a^2 \frac{D C_{, \phi}}{C h} \left( 1 - \gamma^{-2} \right) \\
% &+ a^2 \frac{D_{, \phi}}{2 h} \left( 1 - \gamma^{-2} \right) - a^2 \frac{D}{2} \frac{h_{, \phi}}{h^2} \left( 1 + 2 \gamma^{-3} - 3 \gamma^{-2} \right), \nonumber &
% \end{flalign}

\begin{flalign}
\mathcal{Q}_1 = & \frac{1}{2} a^2 C_{,\phi} \left( 1 - 3 \frac{\delta p_c}{\delta \rho_c} \right) - 3 D \mathcal{H}  \frac{\delta p_c}{\delta \rho_c}  \phi' - \frac{D C_{, \phi}}{C} \phi'^2 +  \frac{D_{, \phi}}{2 }  \phi'^2  - a^2 \frac{D}{2} \frac{h_{, \phi}}{h^2} \left( 1 - \frac{3}{2} \gamma^{-2} \right) + a^2 D \frac{Q}{h \rho_c} \gamma^{-2}, &
\end{flalign}

\begin{flalign}
\mathcal{Q}_2 = &  - \frac{D}{2} \gamma^{-2} \left( 1 + w_c \right) \phi', &
\end{flalign}

\begin{flalign}
\mathcal{Q}_3 = &  3D \mathcal{H} \left( 2 - 3 \gamma^{-2} - w_c \right) - 2 \frac{D C_{, \phi}}{C} \phi' + D_{, \phi} \phi' - 3 D \frac{h_{, \phi}}{h} \left( 1 - \gamma^{-1} \right) \phi' + 3 D \mathcal{H} \gamma^{-1} \left( V_{, \phi} + Q \right) \phi' + 2 D \frac{Q}{\rho_c} \phi', &
\end{flalign}

\begin{flalign}
\mathcal{Q}_4 = &  - k^2 D \left( \gamma^{-2} + w_c \right)  + 3  \mathcal{H}  \left( \frac{D C_{, \phi}}{C} - D_{, \phi} \right) w_c  \phi' - \frac{a^2}{2} \frac{C_{, \phi}^2}{C} \left( 1 - 3 w_c \right) + 2  \frac{D C_{, \phi}}{C }  \phi'^2 -  \frac{3}{2} \frac{C_{, \phi} D_{, \phi}}{C } \phi'^2   \nonumber  \\
& - \frac{3}{2} D \frac{h_{, \phi}}{h} \mathcal{H} \left( 1 - \gamma^{-2} \right) \phi'  + \frac{a^2}{2} \frac{D C_{, \phi}}{C} \frac{h_{, \phi}}{h^2} \left( 1 - \frac{3}{2} \gamma^{-2} \right) - \frac{a^2}{2} D_{, \phi} \frac{h_{, \phi}}{h^2} \left( 1 - \frac{3}{2} \gamma^{-2}  \right)   \nonumber \\
& + \frac{a^2}{2} D \frac{h_{, \phi}^2}{h^3} \left( \frac{5}{4}  - \frac{21}{4} \gamma^{-2} + 4\gamma^{-3} \right) + \frac{1}{2} a^2 C_{, \phi \phi} \left(1 - 3 w_c \right) -  \frac{D C_{, \phi \phi}}{C } \phi'^2 + \frac{D_{, \phi \phi}}{2}  \phi'^2  \\
& - \frac{a^2}{2} D \frac{h_{, \phi \phi}}{h^2} \left( 1 -3 \gamma^{-2} + 2 \gamma^{-3} \right) - a^2 D \gamma^{-3} V_{, \phi \phi}  +  \frac{Q}{h \rho_c} \left[a^2 D_{, \phi} -  a^2\frac{D C_{, \phi}}{C } - \frac{3}{2} D h_{, \phi} \phi'^2 \right], &
\end{flalign}

\noindent for general conformal and disformal coupling functions. If we now take $C(\phi) = \left( T_3 h(\phi) \right)^{-1/2} $ and $D(\phi) = \left( h(\phi) / T_3 \right)^{1/2}$, then we have

\begin{equation}
\delta Q = \frac{a^{-2} \rho_c}{\gamma^{-2} + h \rho_c \gamma^{-3} }\left( \mathcal{Q}_1 \delta_c + \mathcal{Q}_2 \mathpzc{h}' + \mathcal{Q}_3 \delta \phi' + \mathcal{Q}_4 \delta \phi \right),
\end{equation}

\noindent where

% CHOOSE!!

% \begin{flalign}
% \mathcal{B}_1 = & -3 h \mathcal{H} \left( \frac{\delta p_c}{\delta \rho_c} + \gamma^{-2} \right) \phi' - a^2 h \left( V_{,\phi} + Q \right) \gamma^{-3} + a^2 \frac{h_{,\phi}}{4h} \left( 3 \gamma^{-2} - 4 \gamma^{-3} + 3 \frac{\delta p_c}{\delta \rho_c} \right), &
% \end{flalign}

\begin{flalign}
\mathcal{Q}_1 = & a^2 \frac{Q}{\rho_c}\gamma^{-2} + 3 h \frac{\delta p_c}{\delta \rho_c} \left(  a^2 \frac{h_{,\phi}}{4h^2} - \mathcal{H} \phi' \right), &
\end{flalign}

\begin{flalign}
\mathcal{Q}_2 = & -\frac{h}{2} \left( \gamma^{-2} + w \right) \phi', &
\end{flalign}

\begin{flalign}
\mathcal{Q}_3 = & 3 h \mathcal{H} \left( 2 - 3 \gamma^{-2} - w \right) + 3 h^2 \left( V_{,\phi} + Q \right) \gamma^{-1} \phi' + 2 h \frac{Q}{\rho_c} \phi' - \frac{3}{2} h_{,\phi} \left( 1 - 2 \gamma^{-1} \right) \phi', &
\end{flalign}

% CHOOSE!!

% \begin{flalign}
% \mathcal{B}_4 = & -k^2 h \left( \gamma^{-2} + w \right) - a^2 \frac{h_{,\phi}^2}{2 h^2} \left( \frac{3}{2} + \frac{3}{2} w - 3 \gamma^{-1} + \gamma^{-3} \right) + a^2 \frac{3}{4} \frac{h_{,\phi \phi}}{h} \left( \gamma^{-2} - \frac{4}{3} \gamma^{-3} + w \right) \\
% &+ 3 h_{,\phi} \mathcal{H} \left( 1 - 2 \gamma^{-2} -w \right) \phi' + a^2 \frac{h_{,\phi}}{h} \frac{Q}{\rho_c} \left( 1 - \gamma^{-2} \right) + a^2 \frac{h_{,\phi}}{2} \left( V_{,\phi} + Q \right) \left( 3 \gamma^{-1} - 5 \gamma^{-3} \right) - a^2 h V_{,\phi \phi} \gamma^{-3}. \nonumber &
% \end{flalign}

\begin{flalign}
\mathcal{Q}_4 = & -k^2 h \left( \gamma^{-2} + w \right) + a^2 \frac{h_{,\phi}^2}{2 h^2} \left( \frac{3}{4} -\frac{15}{4} \gamma^{-2}  + 4 \gamma^{-3} - \frac{3}{2} w \right) + a^2 \frac{3}{4} \frac{h_{,\phi \phi}}{h} \left( \gamma^{-2} - \frac{4}{3} \gamma^{-3} + w \right)  \\
&  - \frac{3}{2} h_{,\phi} \mathcal{H} \left( 1 - \gamma^{-2} +2w \right) \phi' - a^2 h V_{,\phi \phi} \gamma^{-3} - a^2 \frac{h_{,\phi}}{2h} \frac{Q}{\rho_c} \left( 1 - 3 \gamma^{-2} \right). \nonumber &
\end{flalign}

\section{Generic coupling in Newtonian Gauge} \label{app:newt}

In this Appendix, we provide the general expression of the perturbed disformal coupling $\delta Q$, as discussed in Sec. \ref{sec:pert}, in the Newtonian gauge. It is given by 

\begin{equation}
\delta Q = \frac{a^{-2} \rho_c}{ C - \frac{D}{h} \left(1- \gamma^{-2} \right) + D \rho_c \gamma^{-3} } \left( \mathcal{Q}_1 \delta_c + \mathcal{Q}_2 \Phi' + \mathcal{Q}_3 \Psi + \mathcal{Q}_4 \delta \phi' + \mathcal{Q}_5 \delta \phi \right),
\label{dq}
\end{equation}

\noindent with

\begin{flalign}
\mathcal{Q}_1 = & \frac{1}{2} a^2 C_{,\phi} \left( 1 - 3 \frac{\delta p_c}{\delta \rho_c} \right) - 3 D \mathcal{H}  \frac{\delta p_c}{\delta \rho_c}  \phi' - \frac{D C_{, \phi}}{C} \phi'^2 +  \frac{D_{, \phi}}{2 }  \phi'^2  - a^2 \frac{D}{2} \frac{h_{, \phi}}{h^2} \left( 1 - \frac{3}{2} \gamma^{-2} \right) + a^2 D \frac{Q}{h \rho_c} \gamma^{-2}, &
\end{flalign}

\begin{flalign}
&\mathcal{Q}_2 = 3 D \left( \gamma^{-2} + w \right) \phi', &
\end{flalign}

% \begin{flalign}
% \mathcal{A}_3 = &- 6 D \mathcal{H} \left( 1 - 2 \gamma^{-2} - w \right)  \phi'- 3 a^2 D \left( V_{,\phi} + Q \right)  \left( \gamma^{-1} - \gamma^{-3} \right) - 2 a^2 \frac{D}{h} \frac{Q}{\rho_c} \left( 1 - \gamma^{-2} \right) - a^2 \frac{D_{,\phi}}{h} \left( 1 - \gamma^{-2} \right) \nonumber && \\
% &+ 3 a^2 D \frac{h_{,\phi}}{h^2} \left( 1 - \gamma^{-1} - \gamma^{-2} + \gamma^{-3} \right) + 2 a^2 \frac{D}{h} \frac{C_{,\phi}}{C} \left( 1 - \gamma^{-2} \right), &
% \end{flalign}

\begin{flalign}
\mathcal{Q}_3 = &3 D \mathcal{H} \left( 1 +  \gamma^{-2} + 2w \right)  \phi' - a^2 \frac{D_{,\phi}}{h} \left( 1 - \gamma^{-2} \right) + \frac{3}{4} a^2 D \frac{h_{,\phi}}{h^2} \left( 1 - \gamma^{-2} \right) + 2 a^2 \frac{D}{h} \frac{C_{,\phi}}{C} \left( 1 - \gamma^{-2} \right) \nonumber && \\
&+ a^2 \frac{D}{h} \frac{Q}{\rho_c} \left( 1 - \gamma^{-2} \right), &
\end{flalign}

\begin{flalign}
\mathcal{Q}_4 = &  3D \mathcal{H} \left( 2 - 3 \gamma^{-2} - w_c \right) - 2 \frac{D C_{, \phi}}{C} \phi' + D_{, \phi} \phi' - 3 D \frac{h_{, \phi}}{h} \left( 1 - \gamma^{-1} \right) \phi' + 3 D \mathcal{H} \gamma^{-1} \left( V_{, \phi} + Q \right) \phi' + 2 D \frac{Q}{\rho_c} \phi', &
\end{flalign}

\begin{flalign}
\mathcal{Q}_5 = &  - k^2 D \left( \gamma^{-2} + w_c \right)  + 3  \mathcal{H}  \left( \frac{D C_{, \phi}}{C} - D_{, \phi} \right) w_c  \phi' - \frac{a^2}{2} \frac{C_{, \phi}^2}{C} \left( 1 - 3 w_c \right) + 2  \frac{D C_{, \phi}}{C }  \phi'^2 -  \frac{3}{2} \frac{C_{, \phi} D_{, \phi}}{C } \phi'^2   \nonumber  \\
& - \frac{3}{2} D \frac{h_{, \phi}}{h} \mathcal{H} \left( 1 - \gamma^{-2} \right) \phi'  + \frac{a^2}{2} \frac{D C_{, \phi}}{C} \frac{h_{, \phi}}{h^2} \left( 1 - \frac{3}{2} \gamma^{-2} \right) - \frac{a^2}{2} D_{, \phi} \frac{h_{, \phi}}{h^2} \left( 1 - \frac{3}{2} \gamma^{-2}  \right)   \nonumber \\
& + \frac{a^2}{2} D \frac{h_{, \phi}^2}{h^3} \left( \frac{5}{4}  - \frac{21}{4} \gamma^{-2} + 4\gamma^{-3} \right) + \frac{1}{2} a^2 C_{, \phi \phi} \left(1 - 3 w_c \right) -  \frac{D C_{, \phi \phi}}{C } \phi'^2 + \frac{D_{, \phi \phi}}{2}  \phi'^2  \\
&  - \frac{a^2}{2} D \frac{h_{, \phi \phi}}{h^2} \left( 1 -3 \gamma^{-2} + 2 \gamma^{-3} \right) - a^2 D \gamma^{-3} V_{, \phi \phi} +  \frac{Q}{h \rho_c} \left[a^2 D_{, \phi} -  a^2\frac{D C_{, \phi}}{C } - \frac{3}{2} D h_{, \phi} \phi'^2 \right], &
\end{flalign}

\noindent for general conformal and disformal functions, $C(\phi)$ and $D(\phi)$, respectively.

\end{appendices}

% External bibliography database file in the BibTeX format

%\bibliographystyle{apsrev4-1}
\bibliography{darkbib} 

\end{document}